\documentclass[twocolumn]{aastex631}

\usepackage{amsmath}
\usepackage{units}

\usepackage{booktabs}   % This is for the table, but not necessary if the rules are replaces
\usepackage{graphicx}

\graphicspath{{./}{figures/}}

\newcommand{\msun}{\mathrm{M_{\odot}}} % Msun
\newcommand{\Msun}{\ensuremath{\msun}}
\newcommand{\msunyr}{\msun\,{\rm yr}^{-1}} % Msun

\newcommand{\mhe}{M_{\mathrm{He}}} % M He star
\newcommand{\mwd}{M_{\mathrm{WD}}} % M WD
\newcommand{\mch}{M_{\mathrm{Ch}}} % Chandra Mass
\newcommand{\mex}{\ensuremath{M_{\mathrm{Ex}}}} % Exp Mass

\newcommand{\mwdflash}{M^{\mathrm{flash}}_{\mathrm{WD}}}

\newcommand{\mdotup}{\dot{M}_{\mathrm{up}}} % upper stability line
\newcommand{\mdotlow}{\dot{M}_{\mathrm{low}}} % lower stability line
\newcommand{\mdothe}{\dot{M}_{\mathrm{He}}} % Mdot He star
\newcommand{\mdotwd}{\dot{M}_{\mathrm{WD}}} % Mdot WD
\newcommand{\mdotflashlow}{\dot{M}_{\mathrm{min}}} % lower range of He flash
\newcommand{\lsun}{\mathrm{L_{\odot}}} % Lsun
\newcommand{\Lsun}{\ensuremath{\lsun}} % Lsun
\newcommand{\rsun}{\mathrm{R_{\odot}}} % Rsun
\newcommand{\Rsun}{\ensuremath{\rsun}} % Rsun
\newcommand{\Rhe}{R_{\mathrm{He}}} % R He star
\newcommand{\inipara}{(M^{i}_{\mathrm{He}}, M^{i}_{\mathrm{WD}}, \log P^{i}_{\mathrm{d}} )}
\newcommand{\iniMhe}{M^{i}_{\mathrm{He}}}
\newcommand{\iniMwd}{M^{i}_{\mathrm{WD}}}
\newcommand{\iniP}{\log P^{i}_{\mathrm{d}}}

\newcommand{\finalpara}{(M^{f}_{\mathrm{He}}, M^{f}_{\mathrm{WD}}, \log P^{f}_{\mathrm{d}} )}
\newcommand{\finalMhe}{M^{f}_{\mathrm{He}}}

\newcommand{\finalP}{\log P^{f}_{\mathrm{d}}}

\newcommand{\mesa}{{\tt\string MESA}}

\newcommand{\mdotcr}{\dot{M}_{\mathrm{cr}}}

\newcommand{\Kato}{KH04}

\newcommand{\K}{\ensuremath{\mathrm{K}}}
\newcommand{\Teff}{\ensuremath{T_{\mathrm{eff}}}}
\newcommand{\logTeff}{\ensuremath{\log(\Teff/\K)}}
\newcommand{\logL}{\ensuremath{\log(L/\lsun)}}
\newcommand{\logg}{\ensuremath{\log(g/\mathrm{cm\,s}^{-2})}}

\begin{document}

\title{Pre-Explosion Properties of Helium Star Donors to Thermonuclear Supernovae}

\author[0000-0001-9195-7390]{Tin Long Sunny Wong}% \authorcomment2
\affiliation{Department of Astronomy and Astrophysics, University of California, Santa Cruz, CA 95064, USA}
\affiliation{Department of Physics, University of California, Santa Barbara, CA 93106, USA}

\author[0000-0002-4870-8855]{Josiah Schwab}% \authorcomment1
\altaffiliation{Hubble Fellow}
\affiliation{Department of Astronomy and Astrophysics, University of California, Santa Cruz, CA 95064, USA}

\author[0000-0002-6960-6911]{Ylva G\"{o}tberg}% \authorcomment3
\altaffiliation{Hubble Fellow}
\affiliation{The Observatories of the Carnegie Institution for Science,
813 Santa Barbara St., Pasadena, CA 91101, USA}

\correspondingauthor{Tin Long Sunny Wong}
\email{tinlongsunny@ucsb.edu}

\begin{abstract}
  Helium star - carbon-oxygen white dwarf (CO WD) binaries are potential
  single-degenerate progenitor systems of thermonuclear supernovae.
  Revisiting a set of binary evolution calculations using the stellar
  evolution code \texttt{MESA}, we refine our previous predictions
  about which systems can lead to a thermonuclear supernova and then
  characterize the properties of the helium star donor at the time of
  explosion.  We convert these model properties to NUV/optical
  magnitudes assuming a blackbody spectrum and support this approach
  using a matched stellar atmosphere model.  These models will be
  valuable to compare with pre-explosion imaging for future
  supernovae, though we emphasize the observational difficulty of
  detecting extremely blue companions.
  The pre-explosion source detected in association with SN 2012Z has
  been interpreted as a helium star binary containing an initially
  ultra-massive WD in a multi-day orbit.  However, extending our binary
  models to initial CO WD masses of up to $1.2\,\Msun$, we find that
  these systems undergo off-center carbon ignitions and thus are not
  expected to produce thermonuclear supernovae.
  This tension suggests that, if SN 2012Z is associated with a helium
  star - WD binary, then the pre-explosion optical light from the system must be
  significantly modified by the binary environment and/or 
  the WD does not have a carbon-rich interior composition.
\end{abstract} 


\section{Introduction}
\label{sec:intro}

Close helium star (He star) - carbon-oxygen white dwarf (CO WD)
binaries are potential progenitors of thermonuclear supernovae \citep[TN SNe; e.g.,][]{Iben1994}.
For He stars $\approx \unit[1-2]{\msun}$, thermal timescale
mass transfer initiated during their sub-giant or giant phases
occurs at rates $\sim \unit[10^{-6}]{\msunyr}$ such that the
accreted He can be burned in a thermally-stable configuration on the WD \citep[e.g.,][]{Nomoto1982b}.
Because this implies that the WD can efficiently grow, He star - CO WD binaries are a promising
single-degenerate, Chandrasekhar-mass ($\mch$) TN SN
channel \citep{Yoon2003, Wang2009a, Brooks2016, Wang2017i}.%
\footnote{Another important He star - CO WD binary channel, which we do not discuss in this paper, involves
  lower He star masses $\approx \unit[0.4-1.0]{\msun}$.  These binaries have mass transfer rates
  $\sim \unit[3 \times 10^{-8}]{\msunyr}$, which can lead to accumulation of a He shell and its 
  eventual detonation \citep[e.g.,][and references therein]{Neunteufel2019}.}
This has gained popularity as a progenitor channel for Type Iax
supernovae, under the assumption that the CO WD explodes in a pure
deflagration \citep[e.g.,][]{Kromer2013a, Long2014b, Jha2017}.
The necessarily short orbital periods of these binaries
also provide a mechanism for producing the high velocities
of an emerging population of peculiar stellar objects
that may be the partially-disrupted WD remnants of these explosions
\citep{Vennes2017a, Raddi2018b, Raddi2018a, Raddi2019}.

Because this channel invokes the presence of a luminous, evolved He
star donor, pre-explosion imaging of observed TN SNe can provide a
powerful test of the scenario.  A luminous blue point source is
present at the location of the type Iax supernova SN 2012Z in HST
pre-explosion images of its host galaxy NGC 1309 \citep{McCully2014}.
The properties of the pre-explosion source do not allow for an
unambiguous interpretation, but are consistent with He star - CO WD
models from \citet{Liu2010}.  Only one other Type Iax supernova, SN
2014dt in M61, has comparable pre-explosion limits; in that case,
no source was detected \citep{Foley2015a}.

These and future observations motivate theoretical predictions for the
range of He star donor properties expected at around the time of
explosion.
In \citet{Wong2019}, we used stellar evolution calculations that
resolve the stellar structures of both binary components to identify
which He star - CO WD systems (in terms of initial WD mass, the
initial He star mass, and initial orbital period) evolve to form TN
SNe.
These results were broadly consistent with previous work
\citep{Wang2009a, Wang2017i}, but the calculation of the WD evolution
under a self-consistent, time-dependent mass transfer rate allowed
more accurate distinction between the eventual formation of TN SNe
(via central ignition of carbon burning) and alternative outcomes
\citep[via off-center carbon ignition;][]{Brooks2017b,
  Wu2019}.
However, the \citet{Wong2019} models elided a final phase of
thermally-unstable He shell burning (i.e., He novae) that occurs in
many systems.  Since the models did not cover this phase, they did not
provide the He star properties at the time of explosion.  Here, we
revisit these systems with a modified computational approach that
removes this restriction.

% is critical as it allows the models to be observationally tested via
% pre-explosion imaging of SNe (as the donor is generally predicted to
% dominate the optical light from the system).

In Section~\ref{sec:preexp-evol}, we describe our binary stellar
evolution approach, which uses mass retention efficiencies for He
accretion previously calculated by \citet{KH04}.
In Section~\ref{sec:preexp-models}, we characterize the pre-explosion
properties of our He star models.  We compute the magnitudes and
colors of these objects and check the assumption of blackbody emission
against stellar atmosphere calculations.  We compare with the source
detected in coincidence with SN 2012Z \citep{McCully2014} and with the
models of \citet{Liu2010} and \citet{Liu2015b}.  We similarly conclude
that reproducing the source in 2012Z with a He star model appears to
require a massive ($\approx \unit[1.2]{\Msun}$) WD.  In
Section~\ref{sec:massive}, we extend the \citet{Wong2019} models to CO
WDs with this high mass and show that it is particularly hard for such
systems to avoid off-center carbon ignition (meaning they would not
produce TN SNe).  In Section~\ref{sec:conclusions}, we summarize and
conclude.

\section{Pre-explosion Evolution}
\label{sec:preexp-evol}

\begin{figure*}
\gridline{
  \fig{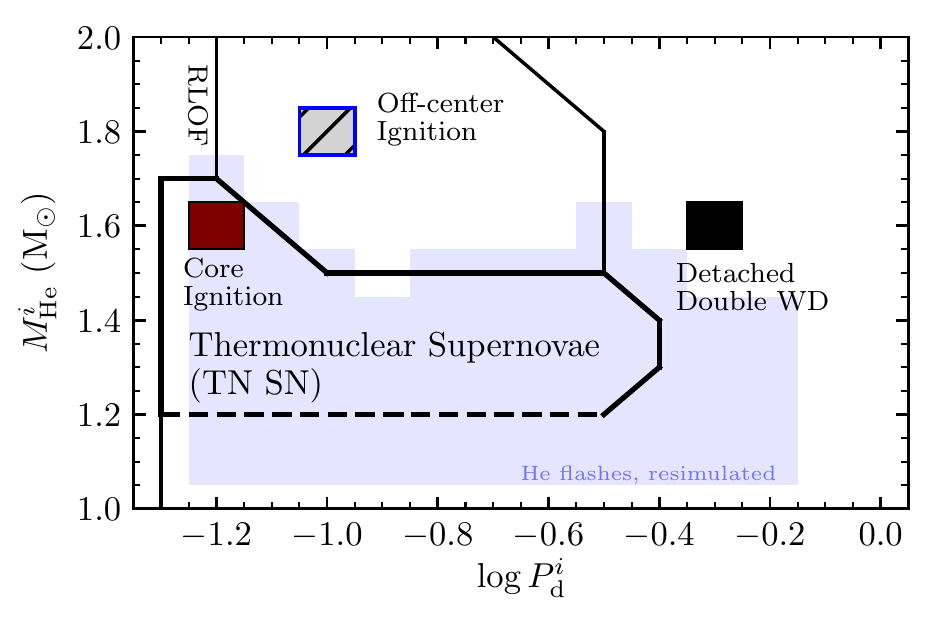}{0.5\textwidth}{(a)}
  \fig{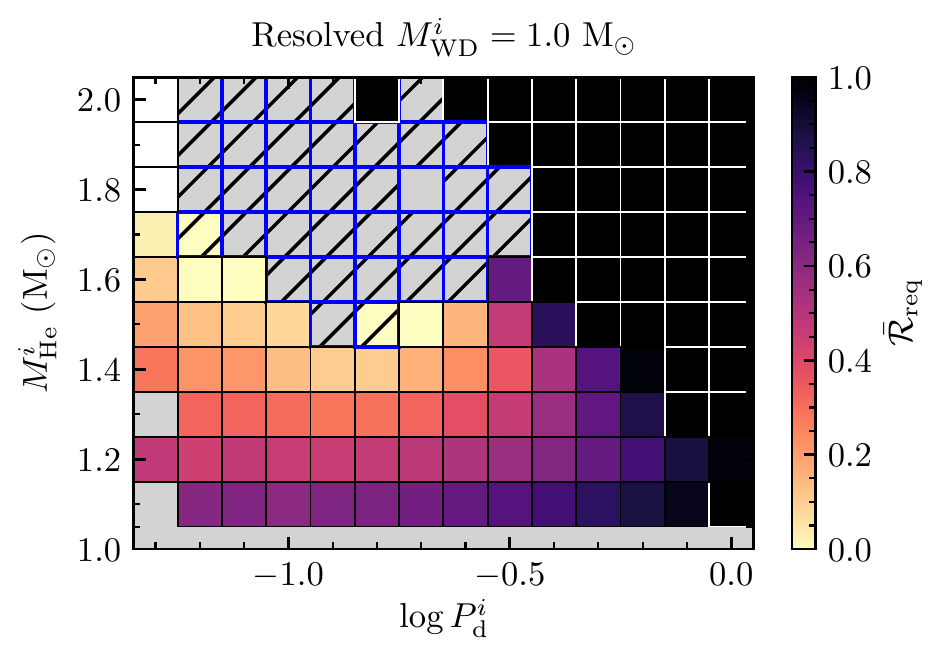}{0.5\textwidth}{(b)}}
\caption{Fate of the He star - CO WD binaries as a function
  of initial He star mass ($\iniMhe$) and initial orbital period ($\iniP$). Panel (a) shows a schematic of the final outcomes of the $\iniMwd = 1.0\,\msun$ grid from \citet{Wong2019}.  Shaded squares label possible outcomes and serve as a key to panel (b).
  The focus of this study is systems in/near the TN SN region, whose lower (dashed) boundary
  is dependent on the assumed average retention efficiency (drawn assuming an efficiency threshold of $\approx 60\%$).  For information on the other outcomes, see \citet{Wong2019}.
  The light shaded background indicates the systems we resimulate in this study. In panel (b), we color code the He flash systems by the required average retention
  efficiency ($\bar{\mathcal{R}}_{\rm req}$) for the WD to grow to $\mch$. 
  Lower efficiencies (redder) are more likely to be TN SNe. Note that some systems at the boundary between outcomes experienced effectively simultaneous core and shell ignitions and thus are marked as both.
  \label{fig:old-grids}}
\end{figure*}

The \citet{Wong2019} He star - CO WD binary models covered a range of
initial CO WD masses ($\iniMwd = 0.90 - 1.05\,\msun$), initial He star
masses ($\iniMhe = 1.1-2.0\,\msun$), and initial orbital periods
($\iniP = -1.3-0.0$).  Computational constraints prevented following
models that experience thermally-unstable He burning through their
many He shell flashes, where the highly super-Eddington conditions in
the envelope lead to prohibitively small timesteps.

Most systems that may eventually grow to $\mch$ experience flashes
during the latter portion of mass transfer.  Therefore, in
\citet{Wong2019}, we stop the binary calculations at the onset of
these flashes and calculate the required average retention efficiency
for the WD to grow up to a critical mass,
$\mex = \unit[1.38]{\Msun} \approx \mch$, the approximate mass for
core carbon ignition to occur in a non-rotating WD.  The required
average retention efficiency is defined as
\begin{equation}
  \mathcal{\bar{R}}_{\mathrm{req}} = \frac{\mex - M^{fs}_{\mathrm{WD}}}{ M^{fs}_{\mathrm{He,env}} },
  \label{eq:reqeff}
\end{equation}
where $M^{fs}_{\mathrm{He,env}}$ is the envelope mass of the He star and $M^{fs}_{\mathrm{WD}}$ is the total mass of the WD, each evaluated when the He flashes first start after thermally-stable mass transfer.

Under the simple but reasonable assumption of a 60\% average retention
efficiency, \citet{Wong2019} classified the final outcome of each
binary system.  Figure~\ref{fig:old-grids}, panel (a) shows a
schematic version of this classification for systems with an initially
$\unit[1.0]{\msun}$ CO WD.  We define the TN SN region as the portion
of parameter space where we observe (or expect) the WD model to
undergo a central carbon ignition as it approaches $\mch$.  The choice
of retention efficiency affects the boundary indicated as a
dashed line.  Figure~\ref{fig:old-grids}, panel (b) shows the required
efficiencies, calculated via Equation~\eqref{eq:reqeff}.

\citet{Wong2019} could approximately
identify which systems reach TN SNe, but could not determine the He
star properties at the time of explosion for the systems that
experience He shell flashes.  This section describes how we resimulate
these systems using point mass accretors in order to determine the
properties of the He stars when the WD explodes.

% Since our previous binary models with resolved WD structures identify
% which systems have off-center ignitions, we re-simulate the binaries
% for which we know there is \emph{not} an off-center ignition using a
% point-mass model for the WD.  By making use of prescriptions for the
% mass retention efficiency during He flashes \citep{KH04, Wu2017}, we
% are able to follow the systems up until the WD reaches $\mch$.  This
% gives us a large set of He star donor models whose properties we know
% at the time of explosion.

\subsection{Methods}
\label{sec:methods}

We simulate the binary evolution of a He star and a point-mass WD accretor using \mesa\ version 10398 \citep{Paxton2011,Paxton2013,Paxton2015,Paxton2018,Paxton2019}, with initial binary parameters ($\iniMhe, \iniMwd$, and $\iniP$) taken from the systems that eventually undergo helium flashes in \cite{Wong2019}.
We adopt the same \mesa\ input options as \citet{Wong2019}, ensuring that the He star evolution should be similar between the sets of calculations.
We briefly recapitulate some of the key choices; for additional details, see \citet{Wong2019} and/or
the publicly available \mesa\ input files.\footnote{\url{https://doi.org/10.5281/zenodo.5540004}}

The He star is created from He ZAMS models assuming $\tt\string Y=0.98$ and $\tt\string Z=0.02$.
We adopt the OPAL Type 2 opacities \citep{iglesias1996} and employ the ``predictive mixing'' capacity of $\mesa$  \citep[for details see Section 2 in][]{Paxton2018} to locate the convective boundary during core He burning.
For numerical convenience we use the \texttt{MLT++} capacity of $\mesa$ to artificially enhance the efficiency of convection in radiation-dominated, near-Eddington conditions.  This is particularly helpful
as the systems begin to come out of contact, when the He star luminosity is highest and its He
envelope is small \citep[see Appendix C in][]{Wong2019}.
For the binary evolution, we adopt the implicit $\tt \string Ritter$ mass transfer scheme.%
\footnote{The assumptions about the structure of the near-Eddington envelope of the He star donor and choice of mass transfer scheme can influence the behavior of the models at the longest orbital periods.
While this can have some qualitative impact on the evolution of individual systems, the
overall conclusions of this paper are not sensitive to this treatment.
Because the purpose of this paper is to demonstrate the effect of
extending the models of \citet{Wong2019}, we continue to adopt
the same treatment of the donor.
This caveat to our models is described more fully in Appendix~\ref{sec:appendixA}.
}
We assume orbital angular momentum loss due to gravitational wave radiation and that systemic mass loss carries the specific angular momentum of the WD accretor.

Point mass models do not evolve the thermal structure of the WD and
therefore require prescriptions to detect the occurrence of off-center
carbon ignition and to handle the fate of the accreted He.  To address
the former, we only re-simulate the binaries from \citet{Wong2019}
where there is not an off-center carbon ignition.  To address the
latter, we adopt existing prescriptions for He accretion in different
regimes, as described in the following subsection and illustrated in Figure~\ref{fig:mdot}.
 
\subsubsection{Mass retention efficiency}

The He star donates material at a rate $\mdothe$, causing the WD to
grow at a net rate $\mdotwd$.  The retention efficiency is the
fraction of mass donated by the He star that is accreted by the WD,
i.e., $\mathcal{R} =
\mdotwd/|\mdothe|$.% and $0 \le \mathcal{R} \le 1$.

When the He star donates material at a rate where the WD
can burn the He in thermally-stable manner, the WD grows at the rate
the material is donated \citep[e.g.,][]{Nomoto1982b, Piersanti2014}.
We indicate the boundaries of this region as $\mdotlow$ and $\mdotup$,
noting that these values are function of $\mwd$ (see Figure \ref{fig:mdot}). The values of $\mdotup$ used in this work are numerically fitted from the models in \cite{Wong2019}.\footnote{This value of $\mdotup$ is slightly smaller than that of \cite{Nomoto1982b}. See Section 5 of \cite{Wong2019} for detailed comparisons to past work.}
(We discuss the values of $\mdotlow$ later.)
We assume that mass transfer is fully conservative ($\mathcal{R} = 1$) 
when the mass transfer rate is $\mdotlow \leq \mdothe \leq \mdotup$.

When the He star donates material faster than the maximum rate at which it can be stably burned, we assume the WD grows at the maximum rate, i.e., $\mdotwd = \mdotup$ for $\mdothe > \mdotup$ and thus $\mathcal{R} = \mdotup/|\mdothe|$ \citep[e.g.,][]{2001ApJ...558..323H,Yoon2003,Wang2009a}. Physically, we assume that material is expelled from the accreting WD either as an isotropic wind or via polar outflows and the material therefore leaves the system with the specific angular momentum of the WD \citep[a scheme also referred to as isotropic re-emission, see][]{1994inbi.conf..263V}. \cite{Wong2019} demonstrated
that which binary systems are predicted to undergo TN SNe is
not sensitive to the assumed specific angular
momentum of this material.

When the He star donates material at a rate below $\mdotlow$, the He-burning shell becomes thermally unstable, leading to He shell flashes that can eject material (i.e., He novae), 
yielding a mass transfer efficiency less than unity.
We use the prescription by \citet{KH04}, hereafter \Kato\, for $\mathcal{R}$ during these flashes. 

The \Kato\ retention efficiencies are based on their optically thick wind theory, where the wind is launched from the iron opacity bump \citep{iglesias1996}. For a fixed WD mass and accretion rate onto the WD, they follow one helium flash cycle by combining a sequence of static and steady-state wind solutions. Over the cycle, the ratio of burned material to ignition mass then gives the retention efficiency. 

We apply the \Kato\ retention efficiencies to our point mass models as follows. These retention efficiencies are only provided for discrete values of the WD mass $\mwdflash$.  For each $\mwdflash$, the efficiency is a function of the accretion rate $\mdotwd$, with an applicable range $\mdotflashlow \leq \mdotwd \leq \mdotlow$.
If $\mdotwd$ is outside the applicable range (i.e., $\mdotwd \leq \mdotflashlow$), we assume a retention efficiency of $\mathcal{R}=0$.
For a given model WD mass $\mwd$, we use the fitting formula of the closest $\mwdflash$ where $\mwdflash \leq \mwd$ (i.e., for a $1.23\,\msun$ WD we adopt the formula for a $1.20\,\msun$ WD). 

We note in passing that He retention efficiencies have also been published by \citet{Piersanti2014} and \citet{Wu2017}. As the former extend only up to a WD mass of $\approx \unit[1]{\msun}$, they cannot be applied in the evolution up to explosion. 
The latter assume that a super-Eddington wind can be driven during a He flash. Their wind mass loss rate is calculated by assuming that the WD luminosity in excess of its Eddington luminosity gives the kinetic power of the wind (assumed to move at the escape velocity). The retention efficiency is then calculated over multiple helium flash cycles in $\mesa$. In general, the \citet{Wu2017} retention effiencies are lower than those of \Kato\ by a factor $\approx 2$. Therefore, our TN SNe region calculated with the \Kato\ prescription would likely be smaller than if calculated with the \citet{Wu2017} prescription.

% % Mass transfer efficiency? 
% \YG{Maybe also here you could refer to Figure~\ref{fig:mdot}? 

% A question: when the white dwarf expels material via a helium flash, does that material also carry the specific angular momentum of the white dwarf? Do helium flashes tighten or widen the binary systems? 
% \sunny{The helium flash also carries the specific angular momentum of the WD. When the helium flashes occur the WD is already more massive than the He star so mass transfer itself tends to widen the orbit; the mass lost due to the helium flashes further widens the orbit. }

% }

In summary, we adopt the prescription
\begin{equation}
\begin{split}
\mdothe >  \mdotup: \quad & \mathcal{R} = \mdotup/|\mdothe| \\
\mdotlow < \mdothe <  \mdotup: \quad & \mathcal{R} = 1 \\
\mdotflashlow < \mdothe <  \mdotlow: \quad & \mathcal{R} = \mathcal{R}_{\rm \Kato} \\
\mdothe <  \mdotflashlow: \quad &
\mathcal{R} = 0 ~.
\end{split}
\end{equation}

~
\subsection{Results}

\begin{figure*}
\fig{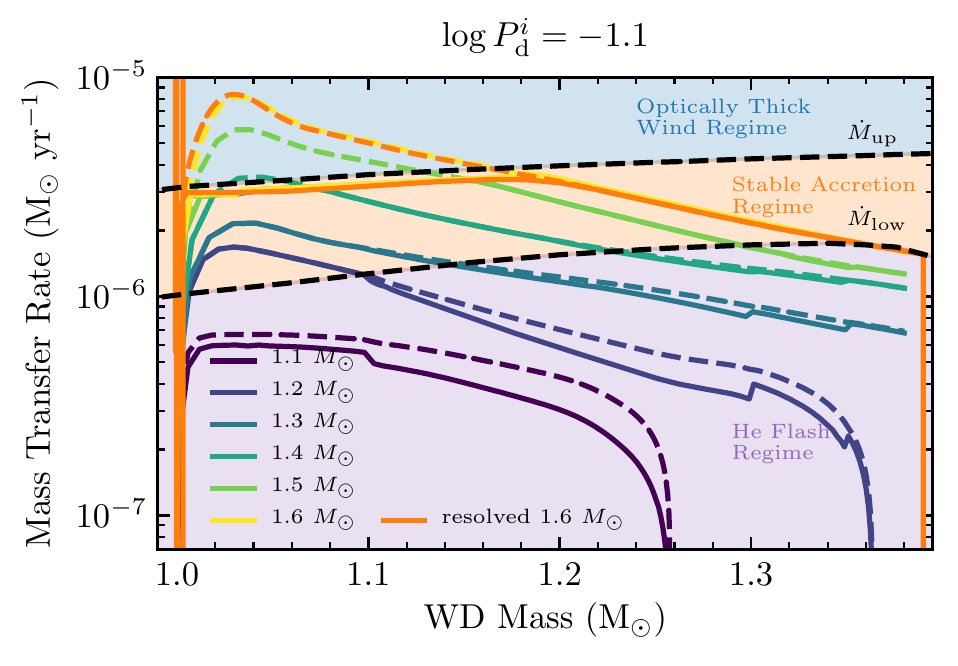}{\textwidth}{}
\caption{Mass transfer histories of the point-mass models for varying $\iniMhe$ with fixed $\iniMwd=1.0 \, \msun$ and $\iniP = -1.1$. The solid lines are $\mdotwd$ and dashed are $|\mdothe|$. The orange lines represent one of the models with a resolved (i.e., non-point-mass) WD. Below $\mdotlow$ we adopt the \Kato \, retention efficiency prescription.
\label{fig:mdot}}
\end{figure*}

We resimulated all of the systems from \citet{Wong2019} that were
halted at the onset of He shell flashes for the $\iniMwd = $ 0.90, 0.95,
1.00, 1.05 $\msun$ WDs.  We summarize the properties of these models
in Table~\ref{tab:models}.

Figure \ref{fig:mdot} shows the mass transfer histories of models at $\iniP=-1.1$. The mass transfer history of the $\iniMhe= 1.6\,\msun$ and resolved WD model (orange lines) is well reproduced by the corresponding point-mass model (yellow lines).  Below $\mdotlow$, the differences between
$\mdotwd$ and $|\mdothe|$ reflect the assumed \Kato\ retention efficiencies.  The 
discontinuities in $\mdotwd$ are due to our piecewise interpolation in these efficiencies.

\startlongtable
\begin{deluxetable*}{c|ccc}
  \tablecolumns{4}
  \tablecaption{We show (from left to right): the initial mass of the helium star, the initial  mass of the white dwarf, the initial period, the final mass of the helium star, the final mass of the white dwarf, the final period, the average helium flash retention efficiency realized in the simulation, and the outcome of the model. The initial values refer to the values before mass transfer is initiated, while the final values refer to the values at the time either the white dwarf explodes or the mass transfer has stopped. If the white dwarf does not experience any helium flashes, the retention efficiency is set to $1$. \label{tab:models}}
  \tablehead{
    \colhead{$\inipara$} & \colhead{$\finalpara$} & \colhead{$\mathcal{\bar{R}}_{\mathrm{sim}}$}  & \colhead{Outcome}
    }
\startdata
(1.1, 0.90, -1.2) & (0.72, 1.19, -1.15) & 0.70 & DWD \\
(1.2, 0.90, -1.2) & (0.75, 1.27, -1.15) & 0.71 & DWD \\
(1.3, 0.90, -1.2) & (0.77, 1.36, -1.16) & 0.81 & DWD \\
(1.4, 0.90, -1.2) & (0.83, \mex, -1.21) & 0.91 & TN SN \\
(1.5, 0.90, -1.2) & (0.89, \mex, -1.24) & 0.95 & TN SN \\
(1.6, 0.90, -1.2) & (0.92, \mex, -1.27) & 0.97 & TN SN \\
(1.7, 0.90, -1.2) & (0.94, \mex, -1.30) & 0.98 & TN SN \\
(1.8, 0.90, -1.2) & (0.96, \mex, -1.34) & 0.98 & TN SN \\
(1.9, 0.90, -1.2) & (0.97, \mex, -1.38) & 0.99 & TN SN \\
(1.1, 0.90, -1.1) & (0.75, 1.18, -1.04) & 0.80 & DWD \\
(1.2, 0.90, -1.1) & (0.78, 1.26, -1.04) & 0.77 & DWD \\
(1.3, 0.90, -1.1) & (0.81, 1.35, -1.07) & 0.87 & DWD \\
(1.4, 0.90, -1.1) & (0.88, \mex, -1.11) & 0.94 & TN SN \\
(1.5, 0.90, -1.1) & (0.95, \mex, -1.15) & 0.98 & TN SN \\
(1.6, 0.90, -1.1) & (0.94, \mex, -1.16) & 0.97 & TN SN \\
(1.7, 0.90, -1.1) & (0.97, \mex, -1.20) & 0.99 & TN SN \\
(1.8, 0.90, -1.1) & (0.98, \mex, -1.23) & 1.00 & TN SN \\
(1.1, 0.90, -1.0) & (0.78, 1.17, -0.95) & 0.86 & DWD \\
(1.2, 0.90, -1.0) & (0.81, 1.25, -0.96) & 0.81 & DWD \\
(1.3, 0.90, -1.0) & (0.83, 1.34, -0.97) & 0.89 & DWD \\
(1.4, 0.90, -1.0) & (0.89, \mex, -1.00) & 0.96 & TN SN \\
(1.5, 0.90, -1.0) & (0.95, \mex, -1.04) & 0.99 & TN SN \\
(1.6, 0.90, -1.0) & (0.98, \mex, -1.07) & 1.00 & TN SN \\
(1.7, 0.90, -1.0) & (0.98, \mex, -1.09) & 1.00 & TN SN \\
(1.8, 0.90, -1.0) & (1.00, 1.37, -1.13) & 1.00 & DWD \\
(1.1, 0.90, -0.9) & (0.81, 1.16, -0.86) & 0.81 & DWD \\
(1.2, 0.90, -0.9) & (0.82, 1.25, -0.86) & 0.84 & DWD \\
(1.3, 0.90, -0.9) & (0.85, 1.32, -0.88) & 0.90 & DWD \\
(1.4, 0.90, -0.9) & (0.88, 1.36, -0.90) & 0.95 & DWD \\
(1.5, 0.90, -0.9) & (0.92, \mex, -0.92) & 0.98 & TN SN \\
(1.6, 0.90, -0.9) & (0.95, 1.37, -0.95) & 0.99 & DWD \\
(1.7, 0.90, -0.9) & (0.99, 1.36, -0.99) & 0.99 & DWD \\
(1.1, 0.90, -0.8) & (0.82, 1.16, -0.77) & 0.82 & DWD \\
(1.2, 0.90, -0.8) & (0.84, 1.24, -0.77) & 0.84 & DWD \\
(1.3, 0.90, -0.8) & (0.87, 1.29, -0.79) & 0.89 & DWD \\
(1.4, 0.90, -0.8) & (0.90, 1.32, -0.80) & 0.93 & DWD \\
(1.5, 0.90, -0.8) & (0.93, 1.33, -0.83) & 0.95 & DWD \\
(1.6, 0.90, -0.8) & (0.96, 1.32, -0.85) & 0.98 & DWD \\
(1.7, 0.90, -0.8) & (1.01, 1.31, -0.89) & 0.99 & DWD \\
\tableline
(1.1, 0.95, -1.2) & (0.73, 1.24, -1.14) & 0.69 & DWD \\
(1.2, 0.95, -1.2) & (0.75, 1.33, -1.14) & 0.76 & DWD \\
(1.3, 0.95, -1.2) & (0.83, \mex, -1.18) & 0.87 & TN SN \\
(1.4, 0.95, -1.2) & (0.91, \mex, -1.23) & 0.95 & TN SN \\
(1.5, 0.95, -1.2) & (0.97, \mex, -1.25) & 0.98 & TN SN \\
(1.6, 0.95, -1.2) & (1.00, \mex, -1.27) & 0.99 & TN SN \\
(1.7, 0.95, -1.2) & (1.03, \mex, -1.30) & 1.00 & TN SN \\
(1.8, 0.95, -1.2) & (1.04, \mex, -1.32) & 1.00 & TN SN \\
(1.9, 0.95, -1.2) & (1.06, \mex, -1.36) & 1.00 & TN SN \\
(2.0, 0.95, -1.2) & (1.06, \mex, -1.39) & 1.00 & TN SN \\
(1.1, 0.95, -1.1) & (0.76, 1.22, -1.03) & 0.78 & DWD \\
(1.2, 0.95, -1.1) & (0.78, 1.31, -1.03) & 0.78 & DWD \\
(1.3, 0.95, -1.1) & (0.84, \mex, -1.07) & 0.90 & TN SN \\
(1.4, 0.95, -1.1) & (0.95, \mex, -1.12) & 0.97 & TN SN \\
(1.5, 0.95, -1.1) & (1.00, \mex, -1.14) & 0.98 & TN SN \\
(1.6, 0.95, -1.1) & (1.03, \mex, -1.16) & 0.99 & TN SN \\
(1.7, 0.95, -1.1) & (1.05, \mex, -1.19) & 1.00 & TN SN \\
(1.8, 0.95, -1.1) & (1.07, \mex, -1.21) & 1.00 & TN SN \\
(1.1, 0.95, -1.0) & (0.78, 1.22, -0.93) & 0.83 & DWD \\
(1.2, 0.95, -1.0) & (0.81, 1.30, -0.94) & 0.90 & DWD \\
(1.3, 0.95, -1.0) & (0.86, \mex, -0.97) & 0.94 & TN SN \\
(1.4, 0.95, -1.0) & (0.95, \mex, -1.01) & 0.98 & TN SN \\
(1.5, 0.95, -1.0) & (1.02, \mex, -1.04) & 1.00 & TN SN \\
(1.6, 0.95, -1.0) & (1.06, \mex, -1.06) & 1.00 & TN SN \\
(1.1, 0.95, -0.9) & (0.81, 1.21, -0.85) & 0.81 & DWD \\
(1.2, 0.95, -0.9) & (0.83, 1.29, -0.84) & 0.83 & DWD \\
(1.3, 0.95, -0.9) & (0.86, 1.38, -0.86) & 0.95 & DWD \\
(1.4, 0.95, -0.9) & (0.94, \mex, -0.90) & 0.99 & TN SN \\
(1.5, 0.95, -0.9) & (1.00, \mex, -0.93) & 1.00 & TN SN \\
(1.6, 0.95, -0.9) & (1.02, \mex, -0.94) & 1.00 & TN SN \\
(1.1, 0.95, -0.8) & (0.82, 1.20, -0.75) & 0.82 & DWD \\
(1.2, 0.95, -0.8) & (0.84, 1.29, -0.75) & 0.85 & DWD \\
(1.3, 0.95, -0.8) & (0.87, 1.35, -0.77) & 0.93 & DWD \\
(1.4, 0.95, -0.8) & (0.91, \mex, -0.78) & 0.98 & TN SN \\
(1.5, 0.95, -0.8) & (0.96, \mex, -0.81) & 1.00 & TN SN \\
(1.6, 0.95, -0.8) & (0.98, \mex, -0.83) & 0.99 & TN SN \\
(1.7, 0.95, -0.8) & (1.01, \mex, -0.85) & 1.00 & TN SN \\
(1.8, 0.95, -0.8) & (1.05, 1.35, -0.88) & 0.95 & DWD \\
(1.9, 0.95, -0.8) & (1.09, 1.29, -0.91) & 0.69 & DWD \\
(1.1, 0.95, -0.7) & (0.83, 1.20, -0.65) & 0.82 & DWD \\
(1.2, 0.95, -0.7) & (0.85, 1.28, -0.65) & 0.86 & DWD \\
(1.3, 0.95, -0.7) & (0.88, 1.33, -0.67) & 0.93 & DWD \\
(1.4, 0.95, -0.7) & (0.91, 1.36, -0.68) & 0.96 & DWD \\
(1.5, 0.95, -0.7) & (0.94, 1.36, -0.70) & 0.99 & DWD \\
(1.6, 0.95, -0.7) & (0.97, 1.35, -0.72) & 0.99 & DWD \\
(1.1, 0.95, -0.6) & (0.84, 1.20, -0.56) & 0.83 & DWD \\
(1.2, 0.95, -0.6) & (0.86, 1.27, -0.56) & 0.87 & DWD \\
(1.3, 0.95, -0.6) & (0.88, 1.31, -0.57) & 0.92 & DWD \\
(1.4, 0.95, -0.6) & (0.91, 1.33, -0.58) & 0.95 & DWD \\
(1.5, 0.95, -0.6) & (0.94, 1.33, -0.60) & 0.97 & DWD \\
(1.1, 0.95, -0.5) & (0.84, 1.19, -0.46) & 0.84 & DWD \\
(1.2, 0.95, -0.5) & (0.86, 1.26, -0.46) & 0.87 & DWD \\
(1.3, 0.95, -0.5) & (0.89, 1.29, -0.47) & 0.90 & DWD \\
(1.4, 0.95, -0.5) & (0.91, 1.30, -0.48) & 0.94 & DWD \\
(1.5, 0.95, -0.5) & (0.95, 1.29, -0.50) & 0.95 & DWD \\
\tableline
(1.1, 1.00, -1.2) & (0.73, 1.28, -1.12) & 0.74 & DWD \\
(1.2, 1.00, -1.2) & (0.77, \mex, -1.14) & 0.84 & TN SN \\
(1.3, 1.00, -1.2) & (0.90, \mex, -1.21) & 0.93 & TN SN \\
(1.4, 1.00, -1.2) & (0.99, \mex, -1.24) & 0.96 & TN SN \\
(1.5, 1.00, -1.2) & (1.04, \mex, -1.26) & 0.99 & TN SN \\
(1.6, 1.00, -1.2) & (1.08, \mex, -1.27) & 1.00 & TN SN \\
(1.7, 1.00, -1.2) & (1.11, \mex, -1.29) & 1.00 & TN SN \\
(1.1, 1.00, -1.1) & (0.76, 1.26, -1.01) & 0.76 & DWD \\
(1.2, 1.00, -1.1) & (0.78, 1.36, -1.01) & 0.84 & DWD \\
(1.3, 1.00, -1.1) & (0.90, \mex, -1.09) & 0.94 & TN SN \\
(1.4, 1.00, -1.1) & (1.01, \mex, -1.13) & 0.98 & TN SN \\
(1.5, 1.00, -1.1) & (1.06, \mex, -1.14) & 0.99 & TN SN \\
(1.6, 1.00, -1.1) & (1.10, \mex, -1.16) & 1.00 & TN SN \\
(1.1, 1.00, -1.0) & (0.78, 1.25, -0.91) & 0.78 & DWD \\
(1.2, 1.00, -1.0) & (0.81, 1.36, -0.92) & 0.87 & DWD \\
(1.3, 1.00, -1.0) & (0.91, \mex, -0.98) & 0.97 & TN SN \\
(1.4, 1.00, -1.0) & (1.01, \mex, -1.02) & 0.99 & TN SN \\
(1.5, 1.00, -1.0) & (1.06, \mex, -1.03) & 1.00 & TN SN \\
(1.1, 1.00, -0.9) & (0.81, 1.25, -0.83) & 0.87 & DWD \\
(1.2, 1.00, -0.9) & (0.83, 1.35, -0.83) & 0.88 & DWD \\
(1.3, 1.00, -0.9) & (0.91, \mex, -0.87) & 0.98 & TN SN \\
(1.4, 1.00, -0.9) & (1.00, \mex, -0.91) & 1.00 & TN SN \\
(1.1, 1.00, -0.8) & (0.82, 1.25, -0.73) & 0.83 & DWD \\
(1.2, 1.00, -0.8) & (0.84, 1.34, -0.74) & 0.89 & DWD \\
(1.3, 1.00, -0.8) & (0.91, \mex, -0.77) & 0.98 & TN SN \\
(1.4, 1.00, -0.8) & (0.98, \mex, -0.79) & 1.00 & TN SN \\
(1.5, 1.00, -0.8) & (1.03, \mex, -0.81) & 1.00 & TN SN \\
(1.1, 1.00, -0.7) & (0.83, 1.25, -0.64) & 0.83 & DWD \\
(1.2, 1.00, -0.7) & (0.85, 1.34, -0.64) & 0.90 & DWD \\
(1.3, 1.00, -0.7) & (0.90, \mex, -0.66) & 0.98 & TN SN \\
(1.4, 1.00, -0.7) & (0.95, \mex, -0.68) & 1.00 & TN SN \\
(1.5, 1.00, -0.7) & (0.99, \mex, -0.69) & 1.00 & TN SN \\
(1.1, 1.00, -0.6) & (0.84, 1.24, -0.54) & 0.84 & DWD \\
(1.2, 1.00, -0.6) & (0.86, 1.33, -0.54) & 0.91 & DWD \\
(1.3, 1.00, -0.6) & (0.89, 1.38, -0.55) & 0.97 & DWD \\
(1.4, 1.00, -0.6) & (0.93, \mex, -0.56) & 1.00 & TN SN \\
(1.5, 1.00, -0.6) & (0.96, \mex, -0.58) & 1.00 & TN SN \\
(1.1, 1.00, -0.5) & (0.84, 1.24, -0.44) & 0.85 & DWD \\
(1.2, 1.00, -0.5) & (0.86, 1.32, -0.44) & 0.91 & DWD \\
(1.3, 1.00, -0.5) & (0.89, 1.36, -0.45) & 0.96 & DWD \\
(1.4, 1.00, -0.5) & (0.91, 1.37, -0.46) & 0.99 & DWD \\
(1.5, 1.00, -0.5) & (0.95, 1.36, -0.47) & 0.99 & DWD \\
(1.6, 1.00, -0.5) & (0.98, 1.34, -0.49) & 0.98 & DWD \\
(1.1, 1.00, -0.4) & (0.84, 1.24, -0.35) & 0.85 & DWD \\
(1.2, 1.00, -0.4) & (0.86, 1.31, -0.34) & 0.91 & DWD \\
(1.3, 1.00, -0.4) & (0.89, 1.34, -0.35) & 0.94 & DWD \\
(1.4, 1.00, -0.4) & (0.92, 1.34, -0.35) & 0.95 & DWD \\
(1.5, 1.00, -0.4) & (0.95, 1.33, -0.37) & 0.97 & DWD \\
(1.1, 1.00, -0.3) & (0.85, 1.24, -0.25) & 0.86 & DWD \\
(1.2, 1.00, -0.3) & (0.87, 1.30, -0.24) & 0.89 & DWD \\
(1.3, 1.00, -0.3) & (0.89, 1.32, -0.25) & 0.94 & DWD \\
(1.4, 1.00, -0.3) & (0.92, 1.32, -0.25) & 0.95 & DWD \\
(1.1, 1.00, -0.2) & (0.85, 1.24, -0.15) & 0.86 & DWD \\
(1.2, 1.00, -0.2) & (0.87, 1.29, -0.15) & 0.89 & DWD \\
(1.3, 1.00, -0.2) & (0.90, 1.30, -0.15) & 0.91 & DWD \\
(1.4, 1.00, -0.2) & (0.92, 1.30, -0.15) & 0.91 & DWD \\
\tableline
(1.1, 1.05, -1.2) & (0.73, 1.33, -1.11) & 0.75 & DWD \\
(1.2, 1.05, -1.2) & (0.84, \mex, -1.18) & 0.88 & TN SN \\
(1.3, 1.05, -1.2) & (0.96, \mex, -1.23) & 0.96 & TN SN \\
(1.4, 1.05, -1.2) & (1.04, \mex, -1.25) & 0.99 & TN SN \\
(1.5, 1.05, -1.2) & (1.10, \mex, -1.26) & 1.00 & TN SN \\
(1.1, 1.05, -1.1) & (0.76, 1.30, -0.99) & 0.73 & DWD \\
(1.2, 1.05, -1.1) & (0.82, \mex, -1.03) & 0.88 & TN SN \\
(1.3, 1.05, -1.1) & (0.96, \mex, -1.09) & 0.94 & TN SN \\
(1.4, 1.05, -1.1) & (1.06, \mex, -1.13) & 0.99 & TN SN \\
(1.5, 1.05, -1.1) & (1.13, \mex, -1.15) & 1.00 & TN SN \\
(1.1, 1.05, -1.0) & (0.78, 1.30, -0.90) & 0.79 & DWD \\
(1.2, 1.05, -1.0) & (0.84, \mex, -0.93) & 0.93 & TN SN \\
(1.3, 1.05, -1.0) & (0.96, \mex, -0.99) & 0.97 & TN SN \\
(1.4, 1.05, -1.0) & (1.07, \mex, -1.02) & 1.00 & TN SN \\
(1.1, 1.05, -0.9) & (0.81, 1.30, -0.81) & 0.85 & DWD \\
(1.2, 1.05, -0.9) & (0.86, \mex, -0.84) & 0.94 & TN SN \\
(1.3, 1.05, -0.9) & (0.97, \mex, -0.88) & 0.99 & TN SN \\
(1.4, 1.05, -0.9) & (1.06, \mex, -0.91) & 1.00 & TN SN \\
(1.1, 1.05, -0.8) & (0.82, 1.30, -0.72) & 0.84 & DWD \\
(1.2, 1.05, -0.8) & (0.86, \mex, -0.73) & 0.96 & TN SN \\
(1.3, 1.05, -0.8) & (0.97, \mex, -0.78) & 1.00 & TN SN \\
(1.4, 1.05, -0.8) & (1.04, \mex, -0.80) & 1.00 & TN SN \\
(1.1, 1.05, -0.7) & (0.83, 1.30, -0.63) & 0.85 & DWD \\
(1.2, 1.05, -0.7) & (0.86, \mex, -0.63) & 0.96 & TN SN \\
(1.3, 1.05, -0.7) & (0.96, \mex, -0.67) & 1.00 & TN SN \\
(1.1, 1.05, -0.6) & (0.84, 1.29, -0.53) & 0.86 & DWD \\
(1.2, 1.05, -0.6) & (0.87, \mex, -0.53) & 0.97 & TN SN \\
(1.3, 1.05, -0.6) & (0.94, \mex, -0.56) & 1.00 & TN SN \\
(1.1, 1.05, -0.5) & (0.84, 1.29, -0.43) & 0.86 & DWD \\
(1.2, 1.05, -0.5) & (0.86, \mex, -0.43) & 0.97 & TN SN \\
(1.3, 1.05, -0.5) & (0.93, \mex, -0.45) & 1.00 & TN SN \\
(1.1, 1.05, -0.4) & (0.84, 1.29, -0.33) & 0.86 & DWD \\
(1.2, 1.05, -0.4) & (0.86, 1.37, -0.33) & 0.97 & DWD \\
(1.3, 1.05, -0.4) & (0.91, \mex, -0.34) & 1.00 & TN SN \\
(1.4, 1.05, -0.4) & (0.95, \mex, -0.35) & 1.00 & TN SN \\
(1.1, 1.05, -0.3) & (0.85, 1.29, -0.24) & 0.87 & DWD \\
(1.2, 1.05, -0.3) & (0.87, 1.36, -0.23) & 0.96 & DWD \\
(1.3, 1.05, -0.3) & (0.90, \mex, -0.23) & 0.99 & TN SN \\
(1.4, 1.05, -0.3) & (0.93, \mex, -0.23) & 1.00 & TN SN \\
(1.1, 1.05, -0.2) & (0.85, 1.29, -0.14) & 0.88 & DWD \\
(1.2, 1.05, -0.2) & (0.87, 1.35, -0.13) & 0.94 & DWD \\
(1.3, 1.05, -0.2) & (0.90, 1.36, -0.13) & 0.98 & DWD \\
(1.4, 1.05, -0.2) & (0.92, 1.36, -0.13) & 0.99 & DWD \\
(1.1, 1.05, -0.1) & (0.86, 1.28, -0.04) & 0.87 & DWD \\
(1.2, 1.05, -0.1) & (0.87, 1.34, -0.03) & 0.94 & DWD \\
(1.3, 1.05, -0.1) & (0.90, 1.35, -0.03) & 0.94 & DWD \\
(1.4, 1.05, -0.1) & (0.93, 1.34, -0.03) & 0.94 & DWD \\
(1.1, 1.05, 0.0) & (0.86, 1.28, 0.06) & 0.87 & DWD \\
(1.2, 1.05, 0.0) & (0.88, 1.33, 0.07) & 0.93 & DWD \\
(1.3, 1.05, 0.0) & (0.90, 1.33, 0.07) & 0.71 & DWD \\
(1.4, 1.05, 0.0) & (0.93, 1.33, 0.07) & 0.94 & DWD \\
\enddata
\tablecomments{
Table 1 is published in its entirety in the machine-readable format.  A portion is shown here for guidance regarding its form and content.}
\end{deluxetable*}

\begin{figure}
  \centering
  \includegraphics[width=\columnwidth]{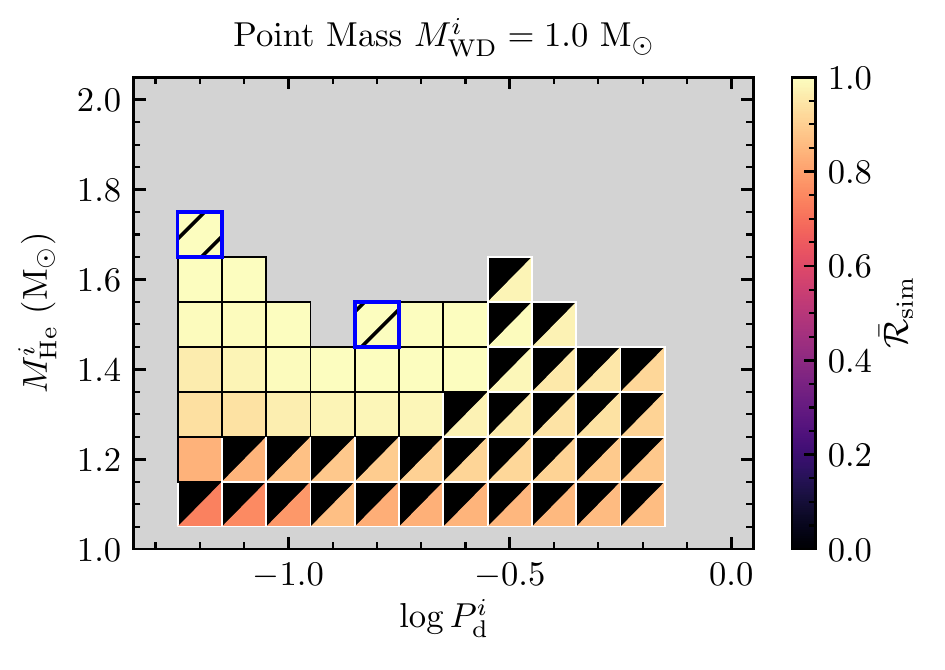}
  \caption{Outcomes of our calculations with point mass accretors. Half-colored points indicate systems that became detached double WDs. Fully colored points reached conditions for a TN SN. Hatched colored points indicate models consistent with either a TN SN or off-center ignition within errors. The color indicates the average He flash retention efficiency realized in the calculation. }
  \label{fig:new-grids}
\end{figure}

The final outcomes of our point-mass WD models are summarized in Figure~\ref{fig:new-grids}. The systems that do reach $\mex$ $(\approx \mch)$ are color coded by the He flash retention efficiency realized in the simulation, defined as:
\begin{equation}
    \mathcal{\bar{R}}_{\mathrm{sim}} = \frac{\mex - M^{fs}_{\mathrm{WD}}}{\mhe^{f} - M^{fs}_{\mathrm{He}}},
\end{equation}
where $M^{fs}_{\mathrm{WD}}$ and $M^{fs}_{\mathrm{He}}$ are the masses of the WD and He star, each evaluated when the He flashes start after thermally-stable mass transfer,
and where $\mhe^{f}$ is the He star mass when the WD reaches $\mex$. For systems that eventually form double detached WDs, we similarly calculate a He flash retention efficiency, but with $\mwd^{f}$ instead of $\mex$.
The systems that we predict to explode have high realized retention fractions, with
$\mathcal{\bar{R}}_{\mathrm{sim}} \gtrsim 0.8$ for the \Kato\ efficiencies. However, this is a necessary but not sufficient condition, as $\mathcal{\bar{R}}_{\mathrm{sim}}$ does not account for the growth of the CO core of the He star, which further depletes the He envelope mass available for mass transfer.

Figure \ref{fig:new-grids} can be compared to panel (b) of Figure \ref{fig:old-grids} which shows the required He flash efficiency $\mathcal{\bar{R}}_{\mathrm{req}}$.
Lower $\iniMhe$ typically leads to higher $\mathcal{\bar{R}}_{\mathrm{req}}$, reflecting the smaller
amount of He available to be donated.
But lower $\iniMhe$ also leads to lower $\mdothe$, and thus lower $\mathcal{\bar{R}}_{\mathrm{sim}}$.
These opposing trends lead to the 
lower limit in $\iniMhe$ to form a $\approx \mch$ WD, occurring where $\mathcal{\bar{R}}_{\mathrm{sim}} \approx \mathcal{\bar{R}}_{\mathrm{req}}$.
Compared to the more approximate result of \citet{Wong2019}, the lower boundary of the TN SN region shifts to slightly higher $\iniMhe$.
%
%Comparing the panels of Figure \ref{fig:new-grids}, we note that a few more systems become detached double WD binaries with the \Wu\ prescription, consistent with the fact that it has a lower retention efficiency than the \Kato\ prescription.  Given the overall similarity between the results, in the remainder of the paper we will consider only the \Kato\ prescription.

\section{The Pre-explosion Models}
\label{sec:preexp-models}

\begin{figure*}
\gridline{
\fig{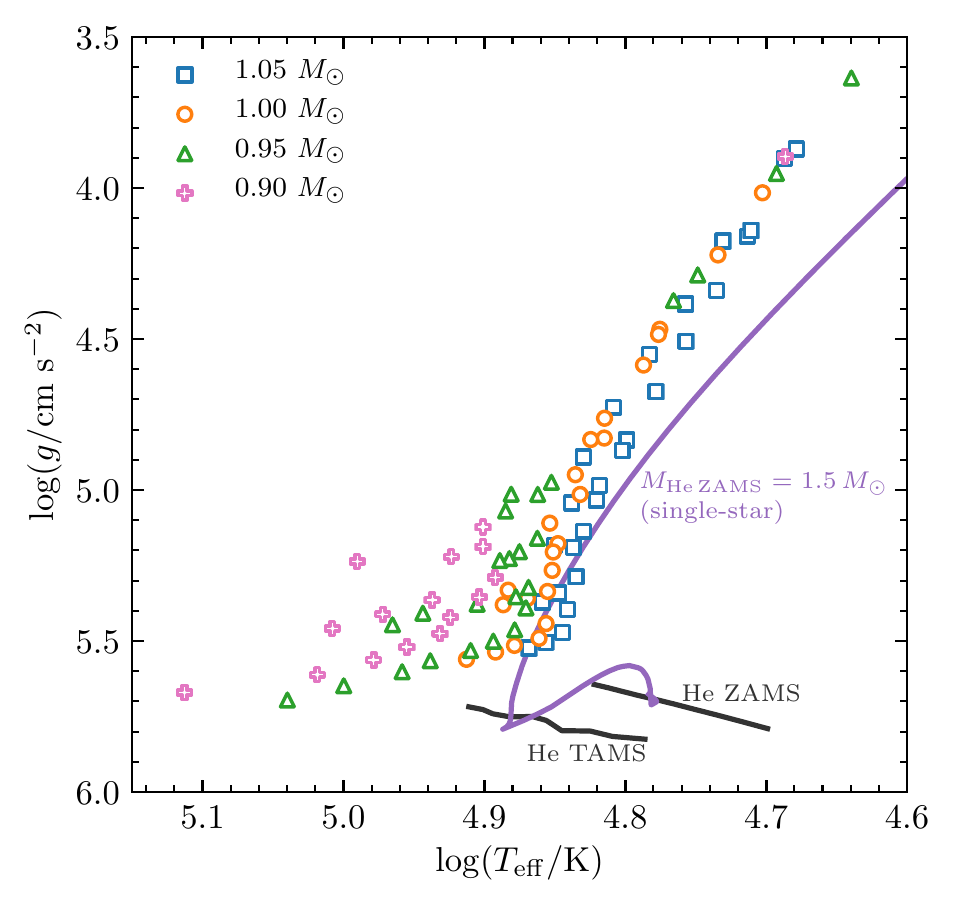}{0.5\textwidth}{(a)}
\fig{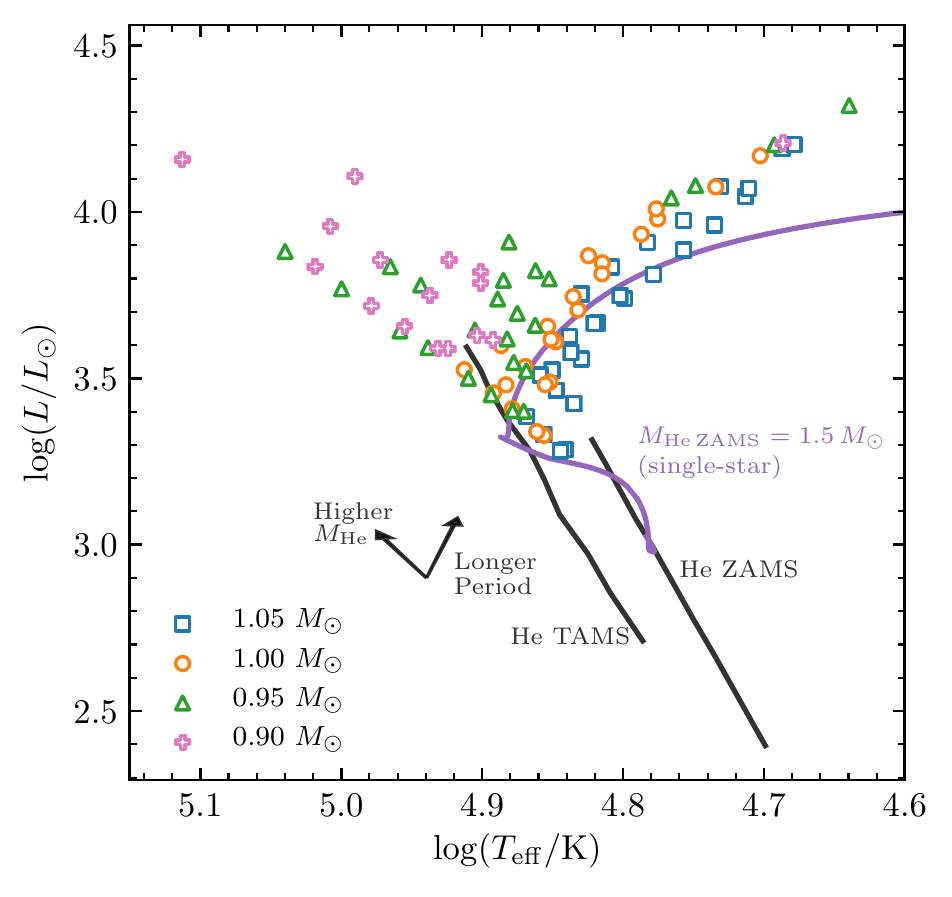}{0.5\textwidth}{(b)}
          }
\caption{
  Kiel diagram (panel a) and Hertzsprung-Russell diagram (panel b) for pre-explosion He star models. Different style points show label different initial WD masses. Wolf-Rayet stars span a similar range in $\log \, T_{\mathrm{eff}}$ and down to $ \logTeff \approx 4.0$, but occupy an area above both plots with lower surface gravity $(\logg \lesssim 4.0)$ and significantly higher luminosity $(\logL \gtrsim 5.0)$. Subdwarfs occupy an area below both plots with similar or higher surface gravity and lower luminosity. 
  \label{fig:logg-logTeff}}
\end{figure*}

\startlongtable
\begin{deluxetable*}{cccccccccccc}
\tablecaption{The properties of the stripped He star companions at the time the white dwarf explodes. We only show the properties of the helium stars in the systems that exploded in a TN SN.\label{tab:he_star_properties}}
\tablehead{
\colhead{$\inipara$} & \colhead{$\finalMhe$} & \colhead{$\log_{10} L$} & \colhead{$\log_{10} T_{\text{eff}}$}  & \colhead{$R$} & \colhead{$\log_{10} g$} & \colhead{$M_{\text{He env}}$}  & \colhead{$M_{\text{F275W}}$} & \colhead{$M_{\text{F336W}}$} & \colhead{$M_{\text{F438W}}$} & \colhead{$M_{\text{F555W}}$}  & \colhead{$M_{\text{F814W}}$} \\
 & \colhead{$[\msun]$} & \colhead{[$\lsun$]} & \colhead{[K]} &  \colhead{[$\rsun$]} & \colhead{[cm s$^{-2}$]} & \colhead{[$\msun$]} & \colhead{[mag]} & \colhead{[mag]} & \colhead{[mag]} & \colhead{[mag]} & \colhead{[mag]}}
\startdata
(1.4, 0.90, -1.2) & 0.83 & 3.59 & 4.92 & 0.29 & 5.42 & 0.18 & 0.15 & 0.54 & 1.03 & 1.40 & 2.25 \\
(1.5, 0.90, -1.2) & 0.89 & 3.59 & 4.93 & 0.29 & 5.48 & 0.21 & 0.19 & 0.59 & 1.08 & 1.45 & 2.30 \\
(1.6, 0.90, -1.2) & 0.92 & 3.66 & 4.95 & 0.28 & 5.52 & 0.20 & 0.18 & 0.58 & 1.07 & 1.45 & 2.30 \\
(1.7, 0.90, -1.2) & 0.94 & 3.72 & 4.98 & 0.27 & 5.56 & 0.18 & 0.19 & 0.59 & 1.09 & 1.47 & 2.32 \\
(1.8, 0.90, -1.2) & 0.96 & 3.84 & 5.02 & 0.25 & 5.61 & 0.14 & 0.16 & 0.57 & 1.07 & 1.46 & 2.31 \\
 (1.9, 0.90, -1.2) & 0.97 & 4.16 & 5.11 & 0.24 & 5.67 & 0.06 & 0.01 & 0.43 & 0.94 & 1.33 & 2.20 \\
 (1.4, 0.90, -1.1) & 0.88 & 3.62 & 4.89 & 0.35 & 5.29 & 0.21 & -0.13 & 0.26 & 0.74 & 1.11 & 1.95 \\
 (1.5, 0.90, -1.1) & 0.95 & 3.63 & 4.90 & 0.34 & 5.35 & 0.26 & -0.09 & 0.30 & 0.78 & 1.16 & 2.00 \\
 (1.6, 0.90, -1.1) & 0.94 & 3.75 & 4.94 & 0.33 & 5.36 & 0.19 & -0.17 & 0.23 & 0.72 & 1.09 & 1.94 \\
 (1.7, 0.90, -1.1) & 0.97 & 3.86 & 4.97 & 0.32 & 5.41 & 0.16 & -0.20 & 0.20 & 0.70 & 1.07 & 1.93 \\
 (1.8, 0.90, -1.1) & 0.98 & 3.96 & 5.01 & 0.31 & 5.46 & 0.13 & -0.21 & 0.20 & 0.70 & 1.08 & 1.93 \\
 (1.4, 0.90, -1.0) & 0.89 & 3.82 & 4.90 & 0.43 & 5.12 & 0.15 & -0.58 & -0.19 & 0.29 & 0.66 & 1.50 \\
 (1.5, 0.90, -1.0) & 0.95 & 3.79 & 4.90 & 0.41 & 5.19 & 0.20 & -0.50 & -0.11 & 0.37 & 0.75 & 1.59 \\
 (1.6, 0.90, -1.0) & 0.98 & 3.86 & 4.92 & 0.40 & 5.22 & 0.19 & -0.53 & -0.13 & 0.35 & 0.73 & 1.57 \\
 (1.7, 0.90, -1.0) & 0.98 & 4.11 & 4.99 & 0.39 & 5.24 & 0.09 & -0.71 & -0.30 & 0.19 & 0.57 & 1.43 \\
 (1.5, 0.90, -0.9) & 0.92 & 4.21 & 4.69 & 1.79 & 3.90 & 0.07 & -2.89 & -2.56 & -2.13 & -1.79 & -0.99 \\
 \tableline
 (1.3, 0.95, -1.2) & 0.83 & 3.40 & 4.87 & 0.30 & 5.39 & 0.25 & 0.26 & 0.65 & 1.12 & 1.49 & 2.33 \\
 (1.4, 0.95, -1.2) & 0.91 & 3.40 & 4.88 & 0.29 & 5.46 & 0.31 & 0.31 & 0.70 & 1.17 & 1.54 & 2.38 \\
 (1.5, 0.95, -1.2) & 0.97 & 3.45 & 4.89 & 0.29 & 5.50 & 0.32 & 0.29 & 0.68 & 1.16 & 1.53 & 2.37 \\
 (1.6, 0.95, -1.2) & 1.00 & 3.50 & 4.91 & 0.28 & 5.53 & 0.32 & 0.27 & 0.67 & 1.15 & 1.52 & 2.36 \\
 (1.7, 0.95, -1.2) & 1.03 & 3.59 & 4.94 & 0.28 & 5.57 & 0.29 & 0.23 & 0.63 & 1.12 & 1.49 & 2.34 \\
 (1.8, 0.95, -1.2) & 1.04 & 3.64 & 4.96 & 0.27 & 5.60 & 0.27 & 0.24 & 0.64 & 1.13 & 1.51 & 2.36 \\
 (1.9, 0.95, -1.2) & 1.06 & 3.77 & 5.00 & 0.26 & 5.65 & 0.22 & 0.20 & 0.61 & 1.11 & 1.49 & 2.34 \\
 (2.0, 0.95, -1.2) & 1.06 & 3.88 & 5.04 & 0.24 & 5.70 & 0.17 & 0.19 & 0.61 & 1.11 & 1.49 & 2.35 \\
 (1.3, 0.95, -1.1) & 0.84 & 3.62 & 4.88 & 0.37 & 5.23 & 0.19 & -0.20 & 0.18 & 0.66 & 1.03 & 1.87 \\
 (1.4, 0.95, -1.1) & 0.95 & 3.52 & 4.87 & 0.35 & 5.32 & 0.32 & -0.05 & 0.33 & 0.81 & 1.18 & 2.01 \\
 (1.5, 0.95, -1.1) & 1.00 & 3.55 & 4.88 & 0.35 & 5.35 & 0.35 & -0.06 & 0.33 & 0.81 & 1.18 & 2.01 \\
 (1.6, 0.95, -1.1) & 1.03 & 3.65 & 4.91 & 0.34 & 5.38 & 0.31 & -0.12 & 0.27 & 0.75 & 1.13 & 1.97 \\
 (1.7, 0.95, -1.1) & 1.05 & 3.78 & 4.94 & 0.34 & 5.41 & 0.25 & -0.20 & 0.20 & 0.68 & 1.06 & 1.91 \\
 (1.8, 0.95, -1.1) & 1.07 & 3.84 & 4.97 & 0.32 & 5.45 & 0.23 & -0.20 & 0.21 & 0.70 & 1.07 & 1.93 \\
 (1.3, 0.95, -1.0) & 0.86 & 3.79 & 4.88 & 0.45 & 5.07 & 0.15 & -0.63 & -0.24 & 0.24 & 0.61 & 1.45 \\
 (1.4, 0.95, -1.0) & 0.95 & 3.66 & 4.86 & 0.43 & 5.16 & 0.27 & -0.44 & -0.05 & 0.42 & 0.79 & 1.62 \\
 (1.5, 0.95, -1.0) & 1.02 & 3.70 & 4.88 & 0.42 & 5.20 & 0.30 & -0.44 & -0.06 & 0.42 & 0.79 & 1.63 \\
 (1.6, 0.95, -1.0) & 1.06 & 3.74 & 4.89 & 0.41 & 5.23 & 0.31 & -0.46 & -0.07 & 0.41 & 0.78 & 1.62 \\
 (1.4, 0.95, -0.9) & 0.94 & 3.80 & 4.85 & 0.52 & 4.97 & 0.21 & -0.85 & -0.47 & 0.01 & 0.37 & 1.20 \\
 (1.5, 0.95, -0.9) & 1.00 & 3.82 & 4.86 & 0.51 & 5.01 & 0.23 & -0.85 & -0.46 & 0.01 & 0.38 & 1.21 \\
 (1.6, 0.95, -0.9) & 1.02 & 3.91 & 4.88 & 0.52 & 5.01 & 0.20 & -0.94 & -0.55 & -0.07 & 0.29 & 1.13 \\
 (1.4, 0.95, -0.8) & 0.91 & 4.04 & 4.77 & 1.03 & 4.37 & 0.11 & -2.00 & -1.64 & -1.19 & -0.84 & -0.02 \\
 (1.5, 0.95, -0.8) & 0.96 & 4.08 & 4.75 & 1.16 & 4.29 & 0.12 & -2.20 & -1.85 & -1.40 & -1.05 & -0.24 \\
 (1.6, 0.95, -0.8) & 0.98 & 4.20 & 4.69 & 1.73 & 3.95 & 0.09 & -2.84 & -2.51 & -2.08 & -1.73 & -0.94 \\
 (1.7, 0.95, -0.8) & 1.01 & 4.32 & 4.64 & 2.54 & 3.64 & 0.06 & -3.45 & -3.13 & -2.72 & -2.38 & -1.60 \\
 \tableline
 (1.2, 1.00, -1.2) & 0.79 & 3.40 & 4.87 & 0.31 & 5.36 & 0.22 & 0.24 & 0.63 & 1.10 & 1.47 & 2.31 \\
 (1.3, 1.00, -1.2) & 0.90 & 3.32 & 4.85 & 0.30 & 5.44 & 0.34 & 0.36 & 0.74 & 1.22 & 1.58 & 2.42 \\
 (1.4, 1.00, -1.2) & 0.98 & 3.36 & 4.87 & 0.30 & 5.48 & 0.37 & 0.33 & 0.72 & 1.19 & 1.56 & 2.39 \\
 (1.5, 1.00, -1.2) & 1.03 & 3.40 & 4.88 & 0.30 & 5.51 & 0.39 & 0.30 & 0.69 & 1.17 & 1.54 & 2.37 \\
 (1.6, 1.00, -1.2) & 1.08 & 3.46 & 4.89 & 0.29 & 5.54 & 0.39 & 0.27 & 0.65 & 1.13 & 1.50 & 2.35 \\
 (1.7, 1.00, -1.2) & 1.10 & 3.53 & 4.91 & 0.29 & 5.56 & 0.37 & 0.23 & 0.62 & 1.10 & 1.48 & 2.32 \\
 (1.2, 1.00, -1.1) & 0.79 & 3.73 & 4.90 & 0.38 & 5.17 & 0.12 & -0.34 & 0.05 & 0.53 & 0.90 & 1.75 \\
 (1.3, 1.00, -1.1) & 0.91 & 3.48 & 4.85 & 0.36 & 5.27 & 0.31 & -0.06 & 0.32 & 0.79 & 1.16 & 1.99 \\
 (1.4, 1.00, -1.1) & 1.00 & 3.48 & 4.85 & 0.36 & 5.33 & 0.39 & -0.04 & 0.34 & 0.82 & 1.18 & 2.02 \\
 (1.5, 1.00, -1.1) & 1.05 & 3.54 & 4.87 & 0.36 & 5.35 & 0.42 & -0.08 & 0.30 & 0.78 & 1.14 & 1.98 \\
 (1.6, 1.00, -1.1) & 1.09 & 3.61 & 4.89 & 0.35 & 5.38 & 0.38 & -0.14 & 0.25 & 0.73 & 1.10 & 1.94 \\
 (1.3, 1.00, -1.0) & 0.91 & 3.59 & 4.84 & 0.44 & 5.11 & 0.27 & -0.42 & -0.04 & 0.43 & 0.79 & 1.62 \\
 (1.4, 1.00, -1.0) & 1.01 & 3.62 & 4.85 & 0.43 & 5.17 & 0.34 & -0.42 & -0.04 & 0.43 & 0.80 & 1.63 \\
 (1.5, 1.00, -1.0) & 1.07 & 3.64 & 4.86 & 0.43 & 5.21 & 0.39 & -0.42 & -0.04 & 0.44 & 0.80 & 1.64 \\
 (1.3, 1.00, -0.9) & 0.92 & 3.69 & 4.82 & 0.52 & 4.96 & 0.25 & -0.75 & -0.38 & 0.09 & 0.45 & 1.28 \\
 (1.4, 1.00, -0.9) & 0.99 & 3.73 & 4.84 & 0.52 & 5.00 & 0.29 & -0.78 & -0.40 & 0.07 & 0.43 & 1.26 \\
 (1.3, 1.00, -0.8) & 0.91 & 3.79 & 4.81 & 0.63 & 4.79 & 0.21 & -1.11 & -0.74 & -0.27 & 0.09 & 0.91 \\
 (1.4, 1.00, -0.8)* & 0.98 & 3.84 & 4.82 & 0.64 & 4.81 & 0.23 & -1.16 & -0.79 & -0.33 & 0.04 & 0.86 \\
 (1.5, 1.00, -0.8) & 1.03 & 3.85 & 4.82 & 0.63 & 4.85 & 0.26 & -1.16 & -0.78 & -0.32 & 0.04 & 0.87 \\
 (1.3, 1.00, -0.7) & 0.90 & 3.90 & 4.79 & 0.80 & 4.58 & 0.16 & -1.53 & -1.17 & -0.71 & -0.35 & 0.46 \\
 (1.4, 1.00, -0.7) & 0.95 & 3.96 & 4.78 & 0.86 & 4.55 & 0.17 & -1.67 & -1.31 & -0.85 & -0.49 & 0.33 \\
 (1.5, 1.00, -0.7) & 0.99 & 3.98 & 4.78 & 0.88 & 4.55 & 0.18 & -1.72 & -1.36 & -0.90 & -0.54 & 0.27 \\
 (1.3, 1.00, -0.6) & 0.88 & 4.06 & 4.74 & 1.19 & 4.24 & 0.10 & -2.21 & -1.86 & -1.41 & -1.06 & -0.25 \\
 (1.4, 1.00, -0.6) & 0.93 & 4.11 & 4.72 & 1.36 & 4.14 & 0.10 & -2.44 & -2.09 & -1.65 & -1.30 & -0.50 \\
 (1.5, 1.00, -0.6) & 0.96 & 4.13 & 4.72 & 1.44 & 4.11 & 0.11 & -2.53 & -2.19 & -1.75 & -1.40 & -0.60 \\
 \tableline
 (1.2, 1.05, -1.2) & 0.84 & 3.29 & 4.84 & 0.30 & 5.40 & 0.32 & 0.36 & 0.74 & 1.21 & 1.58 & 2.41 \\
 (1.3, 1.05, -1.2) & 0.96 & 3.28 & 4.84 & 0.30 & 5.47 & 0.40 & 0.40 & 0.77 & 1.25 & 1.61 & 2.44 \\
 (1.4, 1.05, -1.2) & 1.04 & 3.33 & 4.86 & 0.30 & 5.50 & 0.44 & 0.35 & 0.73 & 1.20 & 1.57 & 2.40 \\
 (1.5, 1.05, -1.2) & 1.10 & 3.39 & 4.87 & 0.30 & 5.52 & 0.46 & 0.29 & 0.67 & 1.15 & 1.52 & 2.35 \\
 (1.2, 1.05, -1.1) & 0.82 & 3.53 & 4.85 & 0.38 & 5.18 & 0.22 & -0.18 & 0.20 & 0.68 & 1.04 & 1.87 \\
 (1.3, 1.05, -1.1) & 0.96 & 3.42 & 4.83 & 0.37 & 5.29 & 0.40 & -0.02 & 0.35 & 0.82 & 1.19 & 2.02 \\
 (1.4, 1.05, -1.1) & 1.06 & 3.46 & 4.85 & 0.36 & 5.34 & 0.45 & -0.04 & 0.34 & 0.81 & 1.17 & 2.01 \\
 (1.5, 1.05, -1.1) & 1.13 & 3.51 & 4.86 & 0.36 & 5.37 & 0.55 & -0.08 & 0.30 & 0.77 & 1.14 & 1.97 \\
 (1.2, 1.05, -1.0) & 0.84 & 3.63 & 4.84 & 0.46 & 5.04 & 0.21 & -0.51 & -0.13 & 0.34 & 0.70 & 1.53 \\
 (1.3, 1.05, -1.0) & 0.96 & 3.56 & 4.83 & 0.44 & 5.14 & 0.34 & -0.39 & -0.02 & 0.45 & 0.82 & 1.64 \\
 (1.4, 1.05, -1.0) & 1.07 & 3.58 & 4.84 & 0.43 & 5.19 & 0.44 & -0.39 & -0.02 & 0.45 & 0.82 & 1.65 \\
 (1.2, 1.05, -0.9) & 0.86 & 3.75 & 4.83 & 0.55 & 4.89 & 0.17 & -0.88 & -0.50 & -0.03 & 0.33 & 1.16 \\
 (1.3, 1.05, -0.9) & 0.97 & 3.67 & 4.82 & 0.52 & 4.98 & 0.30 & -0.73 & -0.36 & 0.10 & 0.47 & 1.29 \\
 (1.4, 1.05, -0.9) & 1.06 & 3.67 & 4.82 & 0.52 & 5.03 & 0.39 & -0.72 & -0.34 & 0.12 & 0.48 & 1.31 \\
 (1.2, 1.05, -0.8) & 0.86 & 3.83 & 4.81 & 0.67 & 4.73 & 0.15 & -1.22 & -0.85 & -0.38 & -0.02 & 0.80 \\
 (1.3, 1.05, -0.8) & 0.97 & 3.74 & 4.80 & 0.62 & 4.83 & 0.27 & -1.04 & -0.67 & -0.21 & 0.15 & 0.97 \\
 (1.4, 1.05, -0.8) & 1.04 & 3.75 & 4.80 & 0.62 & 4.87 & 0.34 & -1.04 & -0.67 & -0.21 & 0.15 & 0.97 \\
 (1.2, 1.05, -0.7) & 0.86 & 3.91 & 4.78 & 0.82 & 4.55 & 0.13 & -1.56 & -1.20 & -0.74 & -0.38 & 0.43 \\
 (1.3, 1.05, -0.7) & 0.96 & 3.81 & 4.78 & 0.75 & 4.67 & 0.24 & -1.35 & -0.99 & -0.53 & -0.17 & 0.64 \\
 (1.2, 1.05, -0.6) & 0.87 & 3.97 & 4.76 & 0.99 & 4.38 & 0.12 & -1.88 & -1.53 & -1.08 & -0.72 & 0.09 \\
 (1.3, 1.05, -0.6) & 0.94 & 3.89 & 4.76 & 0.90 & 4.51 & 0.20 & -1.67 & -1.31 & -0.86 & -0.50 & 0.31 \\
 (1.2, 1.05, -0.5) & 0.86 & 4.08 & 4.73 & 1.26 & 4.17 & 0.08 & -2.30 & -1.95 & -1.51 & -1.16 & -0.35 \\
 (1.3, 1.05, -0.5) & 0.93 & 3.96 & 4.74 & 1.08 & 4.34 & 0.16 & -1.98 & -1.64 & -1.19 & -0.84 & -0.03 \\
 (1.3, 1.05, -0.4) & 0.91 & 4.05 & 4.71 & 1.32 & 4.16 & 0.12 & -2.33 & -1.99 & -1.55 & -1.20 & -0.40 \\
 (1.4, 1.05, -0.4) & 0.95 & 4.07 & 4.71 & 1.37 & 4.14 & 0.13 & -2.41 & -2.07 & -1.63 & -1.28 & -0.48 \\
 (1.3, 1.05, -0.3) & 0.90 & 4.19 & 4.69 & 1.76 & 3.90 & 0.06 & -2.85 & -2.51 & -2.08 & -1.74 & -0.95 \\
 (1.4, 1.05, -0.3) & 0.93 & 4.20 & 4.68 & 1.85 & 3.87 & 0.07 & -2.93 & -2.60 & -2.17 & -1.83 & -1.03 \\
\enddata
\tablecomments{The magnitudes are absolute AB magnitudes and calculated assuming a blackbody spectrum.
Table 2 is published in its entirety in the machine-readable format.  A portion is shown here for guidance regarding its form and content.}
\end{deluxetable*}

For each model in Table~\ref{tab:models} indicated as a TN SN outcome using the \Kato\ retention efficiencies,
Table~\ref{tab:he_star_properties} contains an entry describing the
properties of the He stars at the time of explosion.  They have masses
ranging from 0.75 to 1.15 $\msun$ and remaining He envelope masses
ranging from 0.06 to 0.55 $\msun$.  With increasing $\iniMwd$, $\finalMhe$
is higher with a thicker helium envelope, as less mass is required to
grow the WD up to $\mch$.  Figure~\ref{fig:logg-logTeff} shows the
location of these models in the Kiel and HR diagrams.  Most of the He
stars have $\logg$ between 4.5 and 5.5, while the lowest is around
4.0.  The figure shows that $\logL$ ranges from 3.2 to 4.2 and
$\Teff$ ranges from $\approx 50$ to $100$ kK.  With increasing $\iniMwd$, $g$ and
$\Teff$ are generally lower and $L$ is generally higher.

The pre-explosion He star fills its Roche lobe, so its radius is largely determined by its initial orbital period $\iniP$ since $\iniMhe$ and $\iniMwd$ span a narrow range. As a result, models with the same $\iniP$ lie approximately on the same line of constant radius on the HR diagram (Figure \ref{fig:logg-logTeff}, panel b), with small deviations originating from the $\approx 0.1$ dex change in period due to mass transfer. For models with the same $\iniP$, a higher $\iniMhe$ leads to a higher pre-explosion $L$ and $T_\mathrm{eff}$. On the other hand, with increasing $\iniMwd$, the TN SN region moves to lower $\iniMhe$ and extends to longer $\iniP$ \citep[see Section 4.4,][]{Wong2019}. Therefore, as $\iniMwd$ increases, the pre-explosion models move to lower $L$ and $T_\mathrm{eff}$ with fixed $\iniP$, and to higher $L$ and lower $T_\mathrm{eff}$ with fixed $\iniMhe$.

\citet{Wang2009c} use the results of \citet{Wang2009a} along with
binary population synthesis calculations to predict the donor
properties at explosion.  As discussed in detail in \citet{Wong2019},
our results are in general agreement with their work.  In terms of the
pre-explosion donor properties, one can directly compare our
Figure~\ref{fig:logg-logTeff}a with Figure 2 in \citet{Wang2009c}.  The
models span a similar range of $g$ and \Teff.

%\YG{I would appreciate more descriptions of Figure~\ref{fig:logg-logTeff}. There is a lot of things to say. For example, it is clear that the same initial mass helium star ends up at different locations in the HR diagram, is that because they are all filling their Roche lobes? Which stars actually cause the white dwarf companion to explode and which do not? There appears to be trends for each white dwarf mass, but is there also for each helium star mass? 
%%
%What do you think about adding a few evolutionary tracks in the HRD? Also, the HR diagram is very nice, maybe it is worth to make bigger? Maybe it is worth to roughly mark the location  of WR stars and subdwarfs in the same diagram, so that it is visible that these stars are somewhere in-between? Maybe adding the zero-age helium main sequence and the regular ZAMS? All, just to get a feeling for what kind of objects these stars are. I think there is quite many of them that are hotter than detached stripped stars. If you think it would be useful, you can compare to helium core burning helium stars that are not transferring mass by also plotting the models from \citet{2018A&A...615A..78G}. 
%}

\subsection{Pre-explosion Colors}
\label{sec:preexp-colors}

The colors of these progenitor models in the years leading up to
explosion are of particular interest because of the presence of a
luminous blue point source in the HST pre-explosion image of the type
Iax supernova SN 2012Z \citep{McCully2014}. We focus our spectral modeling on the He star since we expect that the accreting WD would have very high effective temperatures around $\logTeff \approx 5.7 - 6$ \citep[see Figure 12 of][]{Brooks2016}, such that the He star spectrum would dominate in optical wavelengths.

%\YG{Maybe mention this earlier? Like in the introduction? I think it is definitely a good motivation for this work, even though of course it is not the only motivation. Could you maybe mention the colors that were measured for this object in magnitudes? }

\begin{figure*}
  \fig{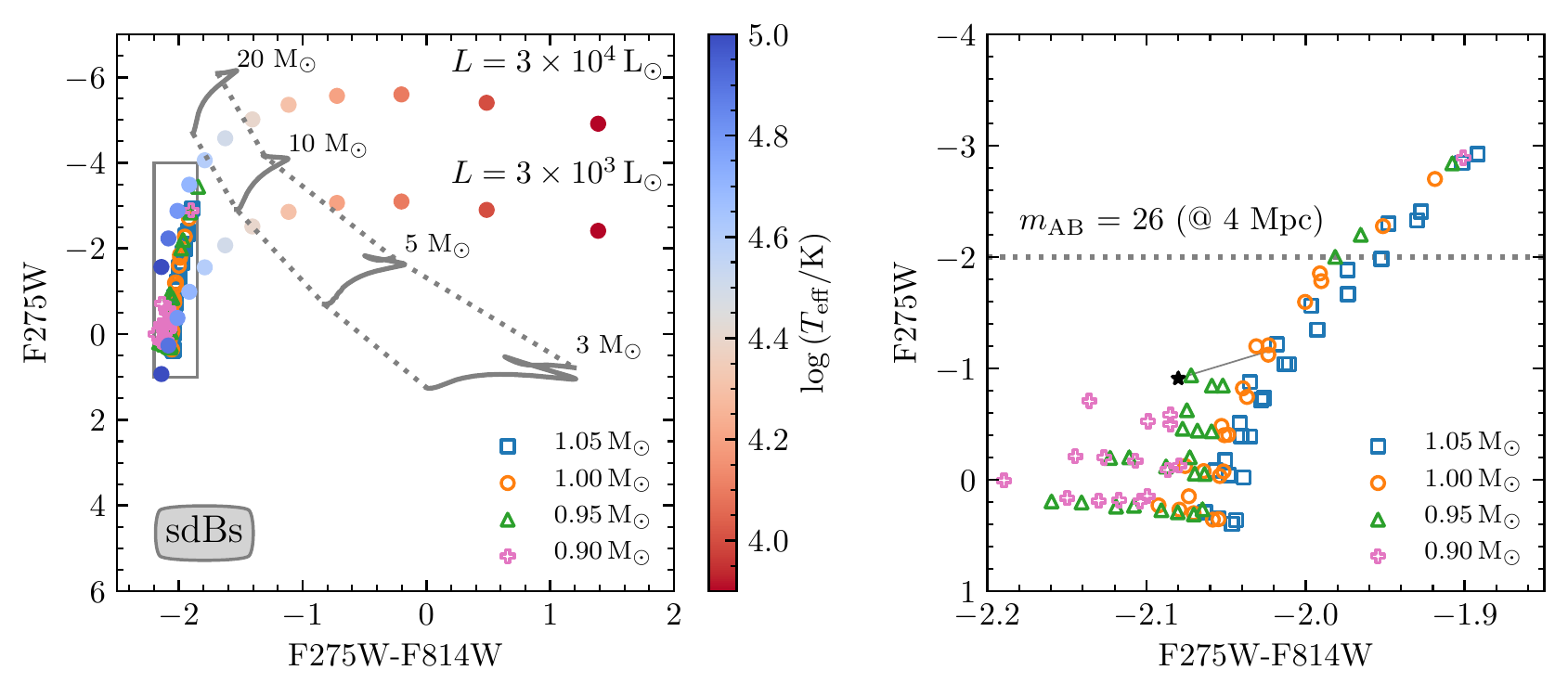}{\textwidth}{}
  %\fig{../colors/WFC3_colors.pdf}{\textwidth}{}
%    \fig{../colors/WFC3_colors_massive.pdf}{\textwidth}{(b)}
%  \includegraphics[width=\textwidth]{../colors/WFC3_colors.pdf}
%  \includegraphics[width=\textwidth]{../colors/WFC3_colors_massive.pdf}
  \caption{Color-magnitude diagrams for indicated WFC3/UVIS filters.
    The pre-explosion He star models shown in
    Figure~\ref{fig:logg-logTeff} are indicated by open symbols.
    These models assume a blackbody spectrum.  In the left panel,
    filled circles show the location of blackbodies with the indicated
    effective temperatures (via the colorbar) and luminosities (via
    the two labeled sequences).  For reference, the main sequence
    (solar metallicity, non-rotating models) from the MIST tracks
    \citep{Choi2016} and the approximate location of sdB stars are
    also shown.  The right panel zooms in on the region indicated by
    the grey rectangle in the left panel.  The black star indicates
    the ``standard'' spectral model from Section~\ref{sec:CMFGEN_test}
    and is connected to its corresponding blackbody model with a thin
    line.}

% }
% \sunny{If we assume a hot SSS WD we might not be able fit it on the CMD (F555W $\ge$ 5). I also want to note that the "SSS" sources in Fig 2 of \cite{McCully2014} probably extend to unreasonably low temperatures (almost $\logTeff \sim 4$). An accretion disk at the rates typical of our simulations coincides with the He stars here. High-mass MS stars fit on this CMD but let's see what happens when we put the accretion disks. }

\label{fig:colors}
\end{figure*}

We use \texttt{starkit} and \texttt{wsynphot} to generate synthetic
photometry for these objects assuming a blackbody
spectrum.\footnote{Available at https://github.com/starkit/}  We report absolute AB magnitudes and colors associated with the HST WFC3/UVIS optical
filters F275W, F336W, F438W, F555W, and F814W. This is motivated by
the Legacy Extragalactic UV Survey (LEGUS), a treasury program that has
observed 50 galaxies within 12 Mpc with this instrument
\citep{Calzetti2015}.  Such data is a potential source of
pre-explosion imaging for future, nearby supernovae.
Figure~\ref{fig:colors} shows color-magnitude diagrams for our
pre-explosion sources.

The luminous helium star donors are sdO stars that sit blueward of
main sequence but well above the sdBs of the extreme horizontal
branch.
They are brightest in the bluest filter F275W with absolute AB
magnitudes of 0.5 to -3 mag and faintest in the red filter F814W with
the absolute magnitudes of 2.5 to -1 mag (see
Table~\ref{tab:he_star_properties}).  Since our pre-explosion models
have $\Teff \gtrsim \unit[50]{k\K}$, the aforementioned filters are
all on the Rayleigh-Jeans tail of the SED (see
Figure~\ref{fig:spectrum}), meaning the colors vary little with
$\Teff$, with a F275W - F814W color of $\approx -2$.

For typical stellar crowding conditions, the LEGUS observations are
designed to reach a depth of $m_{\rm F275W}$ = 26.0 (AB), with
signal-to-noise ratio (S/N) $\sim 6$ \citep{Calzetti2015}.  Since the
LEGUS galaxies are mostly at distances $\approx \unit[4-10]{Mpc}$,
this puts even our coolest models near the edge of detection.
\citet{Calzetti2015} suggest this data set will provide approximately
1 core-collapse SN pre-explosion image per year.  Given local
supernova rates \citep[e.g.,][]{Li2011, Foley2013}, this suggests a
thermonuclear supernova every few years and a decade-scale interval
between SN Iax in this host sample.  The additional extent of the HST
archives makes the situation somewhat less gloomy, but it is clear
that placing limits on extremely blue companions is a significant
challenge.

% \YG{
% Stripped stars are very hot and therefore . We expect that the XX-XX color for most stripped companions is XX mag, while for main sequence stars it would be in the range of XX-XX mag. This difference allows for a clear distinction between a stripped and a main sequence companion and therefore also puts constraints on the evolutionary history if either of the objects were to be discovered prior to a Iax SN explosion"?

% I also wonder whether you could mention detectable color differences? If, for example, the stripped stars have a  0.05 mag bluer color than main sequence stars, can that be distinguished? Maybe, could you also mention which filters that have most archival data and which are best to use for searches of pre-explosion images? 
% }

% The existing pre-explosion source detection is less blue
% than any of the models shown in Figure~\ref{fig:colors}.
%
As discussed in \citet{McCully2014}, the observed source associated
with SN 2012Z is roughly consistent with He star models from
\citet{Liu2010}, who explored the potential of the He
star - WD evolutionary channel to produce super-Chandrasekhar
explosions (via the inclusion of differential rotation in the WD).
Our results for $\approx \mch$ models beginning from a
\unit[1.0]{\Msun} WD are in general agreement with their results
(compare our Figure~\ref{fig:logg-logTeff}, panel (b) with their Figure 7).
However, the \citet{Liu2010} models that are most consistent with the
source in 2012Z are those that result from an initially
\unit[1.2]{\Msun} WD that explodes between \unit[1.4]{\Msun} and
\unit[1.6]{\Msun} (shown in their Figure 6).  These are the
least blue, with $\Teff \sim \unit[10^4]{\K}$, allowing them match the
observed colors and to reach the observed brightness in the optical
bands.

\citet{Liu2015b} perform a similar study using point-mass WD accretors.
Their results for He star - CO WD systems (upper left panel of
their Figure 2) agree with our results in
Figure~\ref{fig:logg-logTeff}, panel (b).  However, in order to match the
properties of the source in 2012Z, they too require a more massive WD
$(\approx 1.2-1.3\,\Msun)$.  Their WD models are point masses, so do
not have a composition.  However, based on difficulties in producing
CO WDs with $\gtrsim \unit{1.1}{\msun}$, they interpret a WD of
this mass to be more likely to be a hybrid CO/ONe WD.%
\footnote{Hybrid CO/ONe WDs have a ONe mantle overlaying an CO core
  and may form
  if mixing at the convective boundary of the inward-going carbon
  flame in a super asymptotic giant branch star quenches burning \citep{Siess2009,
    Denissenkov2013b, Chen2014c}.  There remain questions about
  whether convective mixing can extinguish the flame
  \citep{Lecoanet2016c}.}

% Mixing processes that occur pre-explosion
%   will homogenize the WD \citep{Brooks2017a, Schwab2019}, an effect
%   that has not yet been fully incorporated into the explosion
%   calculations that have explored the potential of such objects to
%   explain SNe Iax \citep{Kromer2015, Bravo2016, Willcox2016}.

Motivated by the indications from these studies that a massive WD
may be needed, we will extend the \citet{Wong2019} models to
higher initial WD masses in Section~\ref{sec:massive}.

% Test with a spectral model to see whether they are needed
\subsection{The predictions from spectral models}\label{sec:CMFGEN_test}

% Table with computed properties of the stripped donor stars
\begin{table*}
\caption{Stellar properties and AB magnitudes predicted by a blackbody and atmosphere models for the stripped He star model $\finalMhe=0.98\, \msun$ at the time the white dwarf explodes.}
\label{tab:CMFGEN_models}
\begin{center}
\begin{tabular}{lccccccccc}
\toprule\midrule
Model & $T_{\text{eff}}$ & $\log_{10} g_{\text{eff}}$ & $v_{\infty}$ & $\dot{M}_{\text{wind}}$ & $M_{\text{F275W}}$ & $M_{\text{F336W}}$ & $M_{\text{F438W}}$ & $M_{\text{F555W}}$ & $M_{\text{F814W}}$ \\ 
 & [kK] & [cm s$^{-2}$] & km s$^{-1}$ & [$M_{\odot}$ yr$^{-1}$] & [mag]& [mag]& [mag]& [mag]& [mag]\\ 
\midrule 
  Blackbody & 65.63 & 4.81 & -- & -- & -1.16 & -0.79 & -0.33 & 0.04 & 0.86 \\
  %$-1.17$ & $-0.8$ & $-0.33$ & $0.02$ & $0.85$ \\ 
\midrule 
standard & 65.27 & 4.8 & 1200 & $10^{-8}$ & $-0.91$  & $-0.51$  & $-0.04$  & $0.31$  & $1.17$ \\ 
slow & 65.27 & 4.8 & 600 & $10^{-8}$ & $-0.91$  & $-0.52$  & $-0.04$  & $0.3$  & $1.16$ \\ 
slower & 65.27 & 4.8 & 300 & $10^{-8}$ & $-0.91$  & $-0.52$  & $-0.04$  & $0.31$  & $1.16$ \\ 
more & 65.27 & 4.8 & 1200 & $10^{-7}$ & $-0.93$  & $-0.59$  & $-0.1$  & $0.19$  & $0.99$ \\ 
extreme & 65.25 & 4.8 & 300 & $10^{-7}$ & $-0.97$  & $-0.66$  & $-0.23$  & $0.02$  & $0.71$ \\ 
\bottomrule
\end{tabular}

\end{center}
\tablecomments{The effective temperature, $T_{\text{eff}}$, and the effective surface gravity, $\log_{10} g_{\text{eff}}$, are predicted from the photosphere at $\tau = 2/3$ in the atmosphere models. The stripped He star has a luminosity of $6.8 \times 10^3 L_{\odot}$, a mass of $0.98 M_{\odot}$, and a radius of $0.64 R_{\odot}$.}
\end{table*}

% Note: Sorry for the delay, the spectral models are now safe to use. There is a bolometric luminosity difference of at max 2\% between the spectral models and the blackbody. This difference could come from re-emission in wavelengths the code is not treating so I don't think it is right to rescale from there. Anyhow, it is a very small difference. 

% \YG{
% TODO: (1) Update the figures, (2) Check that the text makes sense.
% }

% Background -- why bother about spectral models? 
The true emergent spectrum from the photosphere of a He star is not a blackbody and it is therefore possible that the predictions for the pre-explosion photometry change if we account for a more realistic spectral energy distribution for the He stars. Here, we test how accurate the blackbody assumption is by comparing one spectral model computed for one of the stripped He star with its corresponding blackbody spectrum. For this, we use the 1D non-LTE radiative transfer code CMFGEN \citep[][version from 5 May 2017]{1990A&A...231..116H, 1998ApJ...496..407H}.

% What exactly happens here -- method part
We choose to model the spectrum for a 0.98 $\msun$ stripped He star orbiting a $\iniMwd=1.0$ $\msun$ WD on a $\iniP=-0.8$ day orbit at the time the white dwarf explodes (the model is marked with a star in Table~\ref{tab:he_star_properties}. We take the same approach as outlined in \citet{2017A&A...608A..11G} and assume the surface properties computed with \texttt{MESA} as the conditions at the base of the stellar atmosphere \citep[see also][]{2014A&A...564A..30G}. We then model the emerging spectrum after taking assumptions for the wind mass loss rate, the wind speed, and wind clumping. 

Stellar wind mass loss is known to significantly affect the emerging spectrum by, for example, blocking ionizing emission if the wind is optically thick or introducing strong emission features. The wind mass loss from He stars is poorly constrained since very few stars have been observed \citep[see however][]{1998ApJ...493..440G, 2008A&A...485..245G, 2018ApJ...853..156W}. Theoretical predictions suggest that the winds from He stars are weak (with mass loss rates of $\dot{M}_{\text{wind}} \sim 10^{-9} - 10^{-7} \, M_{\odot}$~yr$^{-1}$) and relatively fast (with terminal wind speeds of $v_{\infty} \gtrsim 1000$ km~s$^{-1}$) \citep[e.g.,][]{2016A&A...593A.101K, 2017A&A...607L...8V}. The star we model has a luminosity of $6.8 \times 10^3\,\Lsun$, a mass of $0.98\,\Msun$, and a radius of $0.64\,\Rsun$ at the time the white dwarf exploded\footnote{This model was made during the earlier stages of this work, and so its properties differ slightly from that shown in Table \ref{tab:he_star_properties}.}. We further predicted a surface temperature of 65,300~K and surface gravity of $\logg = 4.8$ from the \texttt{MESA} calculation. Following the theoretical predictions, we assume a wind mass loss rate of $10^{-8} \, M_{\odot}$~yr$^{-1}$ and a terminal wind speed of $1\ 200$~km~s$^{-1}$ for the atmosphere modeling. We assume that the wind follows a non-standard $\beta$-law with wind profile parameter, $\beta$, set to 1, and a somewhat clumpy wind with a volume filling factor of 0.5. For numerical reasons, we included a negligible amount of hydrogen in the atmosphere ($X_{\text{H, s}} = 2.5 \times 10^{-11}$). We refer to this model as the standard model (see Table~\ref{tab:CMFGEN_models}).

% Comparison of spectra
\begin{figure*}
  \centering
  \includegraphics[width=.75\textwidth]{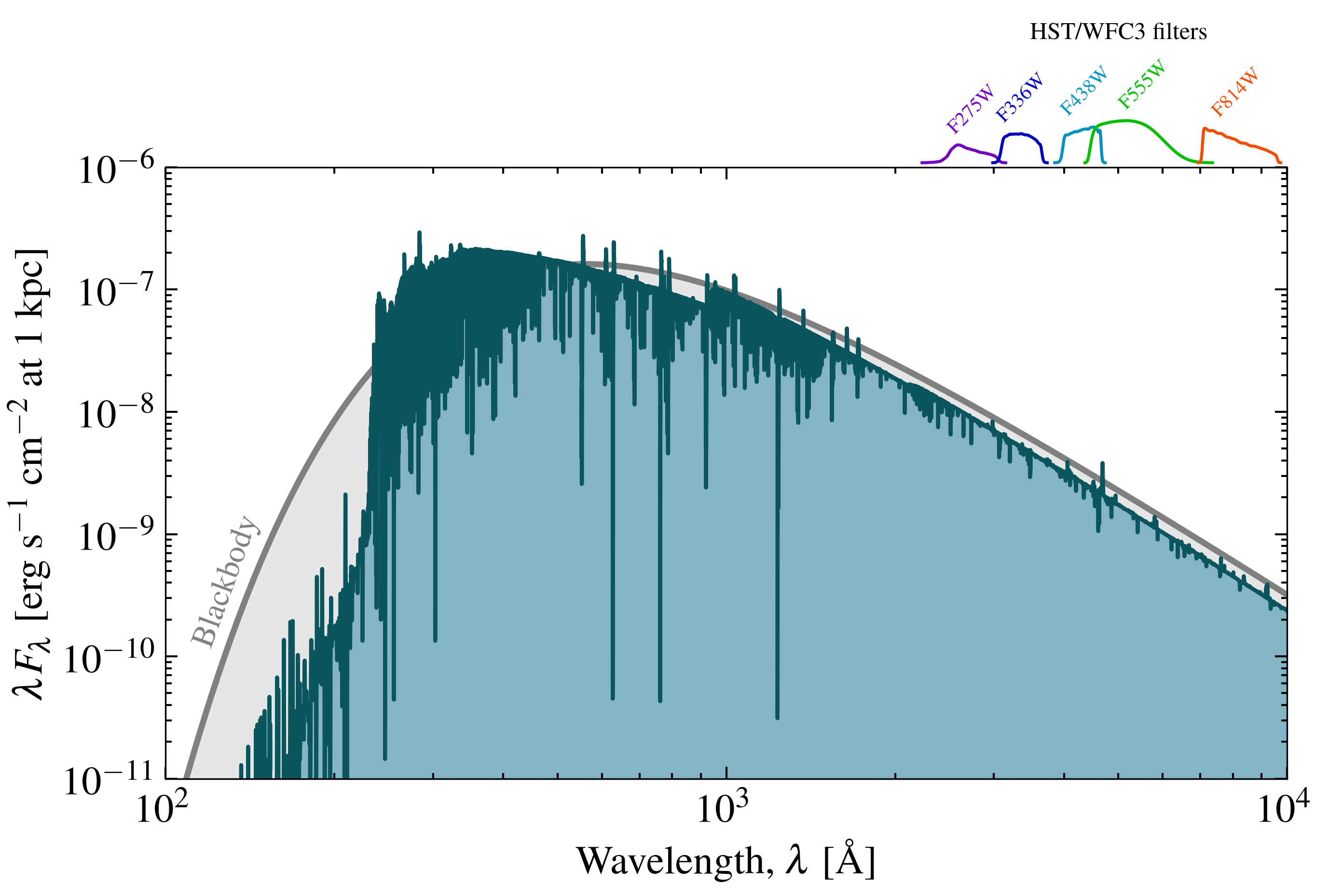}
  \caption{Spectral energy distribution for the stripped He star marked in Table~\ref{tab:he_star_properties} at the time of explosion of the white dwarf companion. We show the predictions from a detailed atmosphere model (standard, see Table~\ref{tab:CMFGEN_models}) together with a blackbody curve. Above the plot, we show the transmission curves for HST/WFC optical filters, which are commonly used when searching for objects in pre-explosion images. Because the He stars are very hot, the filters capture the Rayleigh-Jeans part of the spectrum, in which the difference between the atmosphere model and blackbody curve is relatively small. }
  \label{fig:spectrum}
\end{figure*}

% Results
% Figure shows the SEDs and the magnitudes
We show the resulting spectral energy distribution for the standard model in Figure~\ref{fig:spectrum}. The figure shows that the spectral energy distribution has a similar shape when the spectrum is carefully modeled compared to when a blackbody is assumed. The stellar wind is not sufficiently optically thick to significantly affect the shape of the part of the spectrum that the considered filters probe. The main difference in terms of photometrical estimates is that the spectral model predicts somewhat lower flux in the optical wavelengths compared to the blackbody, corresponding to a systematic difference of about 0.3 mag.
The colors of the modeled spectrum and the blackbody assumption are therefore similar. We present the calculated absolute magnitudes for the HST filters in Table~\ref{tab:CMFGEN_models}. 

% Variations in the wind properties
Since the wind properties of He stars are uncertain, we create four additional models by varying the mass loss rates and/or the wind speed as presented in Table~\ref{tab:CMFGEN_models}. In two models, we decrease the terminal wind speed to 600~km~s$^{-1}$ (slow) and 300~km~s$^{-1}$ (slower). In another model, we increase the wind mass loss rate to $10^{-7}\ \msun$~yr$^{-1}$, but keep the wind speed at $1\ 200$~km~s$^{-1}$ (more). In the last model, we increase the wind mass loss rate to $10^{-7}\ \msun$~yr$^{-1}$ and decrease the terminal wind speed to 300~km~s$^{-1}$ (extreme). With slower winds or higher wind mass loss rates, the wind becomes denser and therefore more optically thick. If the wind is sufficiently dense, the emission at longer wavelengths is enhanced and the color becomes redder. Investigating how the stellar wind affects the spectral energy distribution is therefore important for understanding the origin of objects that are redder than expected, such as the one observed in SN 2012Z. 

\begin{figure}
  \centering
  \includegraphics[width=\columnwidth]{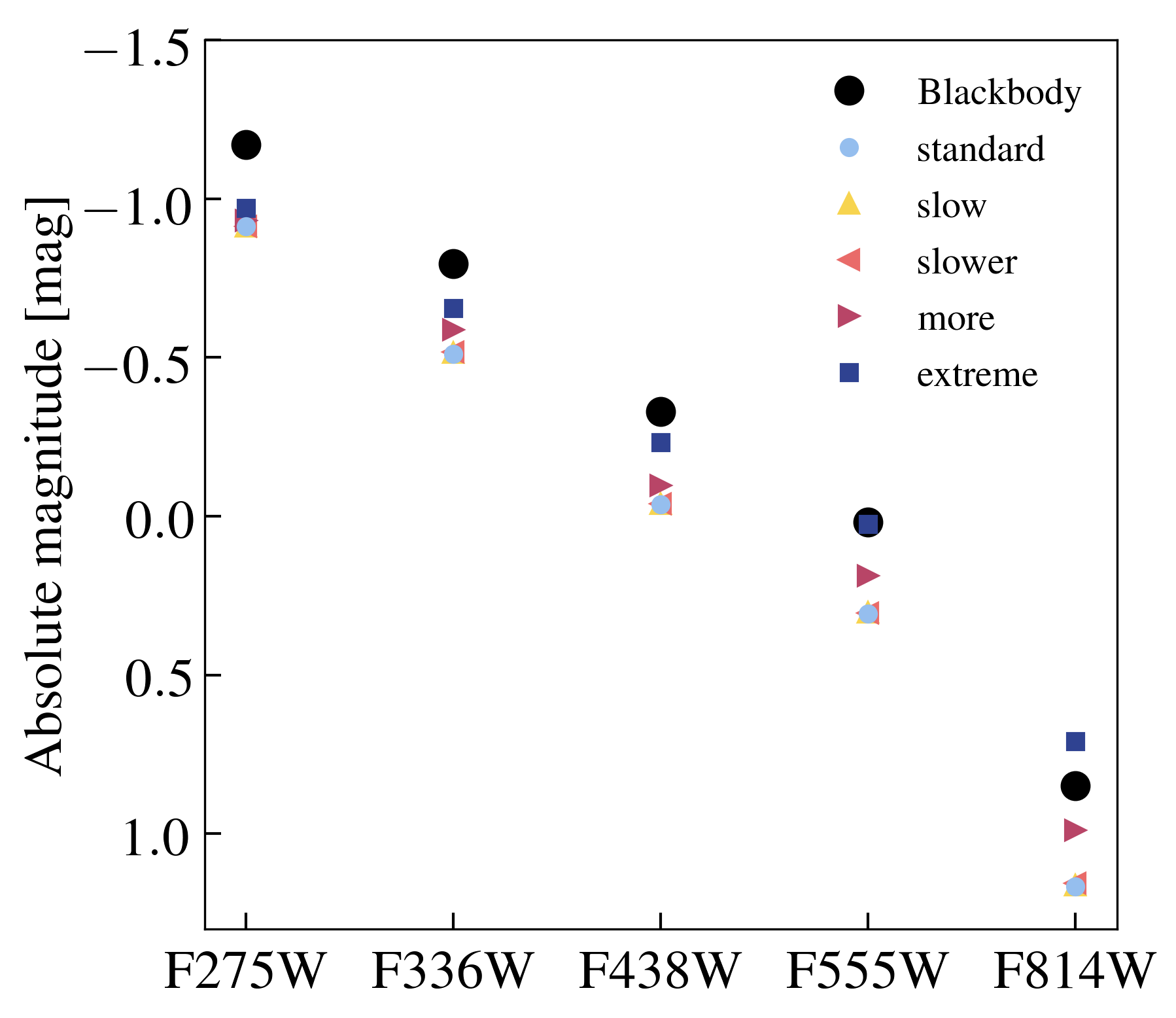}
  \caption{We compare the absolute AB magnitudes predicted from the different spectral models with that of a corresponding blackbody, all computed for the stripped He star marked in Table~\ref{tab:he_star_properties} and discussed in Section~\ref{sec:CMFGEN_test}. In general, the blackbody predicts brighter optical emission than the spectral models, while the predicted colors are similar between the different models.}
%  \caption{\YG{The absolute magnitudes of the HST filters, computed for the stripped companion in the systems XX. We show the predictions from the blackbody assumption in black, that of the faster wind-model in blue, and for the slower wind model in nude color. Include observed colors here?}}
  \label{fig:colors_CMFGEN}
\end{figure}

However, we do not find a large difference in the photometrical magnitudes when varying the wind parameters (see Table~\ref{tab:CMFGEN_models}). The largest difference is seen in the extreme model, with at maximum 0.46 mag difference compared to the standard model in the reddest band, but only 0.06 mag difference in the bluest band. The colors do not significantly change either, F438W$-$F555W is estimated between $-0.25$ mag and $-0.35$ mag, while F555W$-$F814W is estimated between $-0.69$ mag and $-0.86$ mag. This can also be seen in Figure~\ref{fig:colors_CMFGEN} where the predictions for absolute magnitudes are displayed for the blackbody and the spectral models.
Since we created models with large differences in the wind properties compared to what is expected for He stars with the given stellar properties, we can therefore consider that the wind from the He star is not sufficient for making the star as red as observed in SN 2012Z (see also Figure~\ref{fig:colors}).

% Comparison of spectra

\section{Models with Initially Massive WDs}
\label{sec:massive}

In this section, we extend the models of \citet{Wong2019} to include more
massive $\iniMwd=\unit[1.10]{\msun}$ and $\unit[1.20]{\msun}$ CO WDs
and examine the occurrence off-center carbon ignitions.

In \citet{Wang2009c}, when the accretor is a more massive WD, systems
with higher He star masses and longer initial periods are able to
reach explosion (their Figure 8).  Higher He star masses and longer initial periods
lead to increased mass transfer rates, but since the WD is limited
in the amount it can accept, the mass transfer becomes more non-conservative.
This can be counterbalanced by the WD beginning closer to $\mch$.
However, accounting for off-center ignitions
(which eliminates the systems that transfer mass at or above $\mdotup$ for most of their history)
leaves only a narrow range of systems with
$\iniMhe \approx \unit[1.1]{\msun}$ and initial periods out to
\unit[100]{d} \citep[see Figure 8 in][]{Wang2017i}.  The He star
luminosity lies in a narrow range, so in systems with a wider orbit,
the larger Roche lobe allows for a lower effective temperature of the
Roche-lobe-filling He star donor.  A similar line of reasoning leads
\citet{Liu2015b} require a massive WD in a binary with an orbital period $\ga \unit[10]{d}$
in order to match the properties of the pre-explosion source in SN 2012Z.

% \YG{Here, it would be nice if you could clarify why not higher mass He stars are suitable as companions. Also, just a comment for why the longer periods give more non-conservative mass transfer would be helpful. }

\subsection{Point-mass WD binary models}

% Suppose nature manages to produce a massive ($\gtrsim 1.05 \, \msun$) CO WD, then can the CO WD remain so by avoiding an off-center carbon ignition? 

\begin{figure*}
\gridline{
\fig{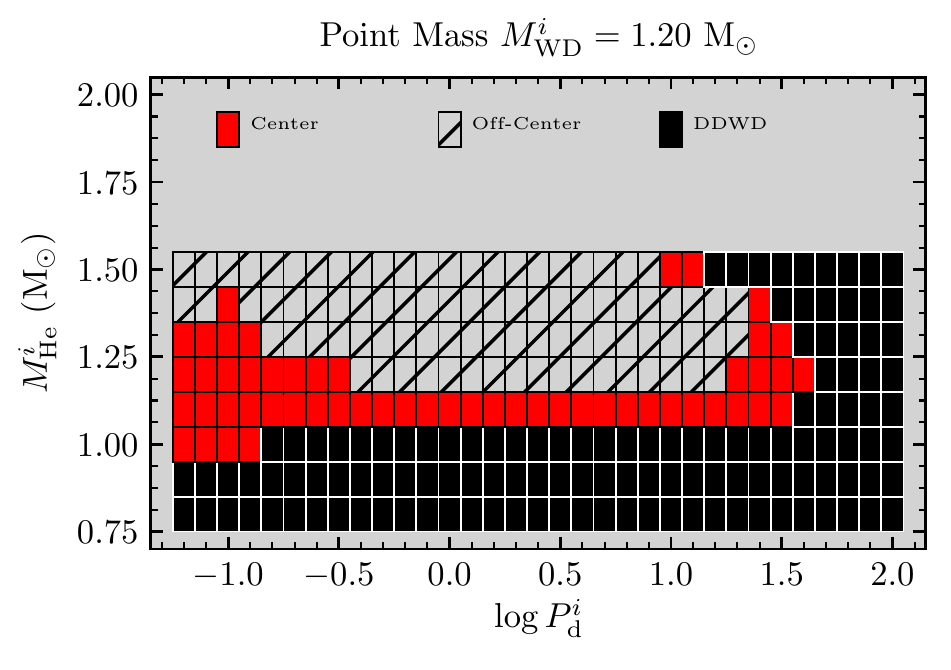}{0.5\textwidth}{(a)}
\fig{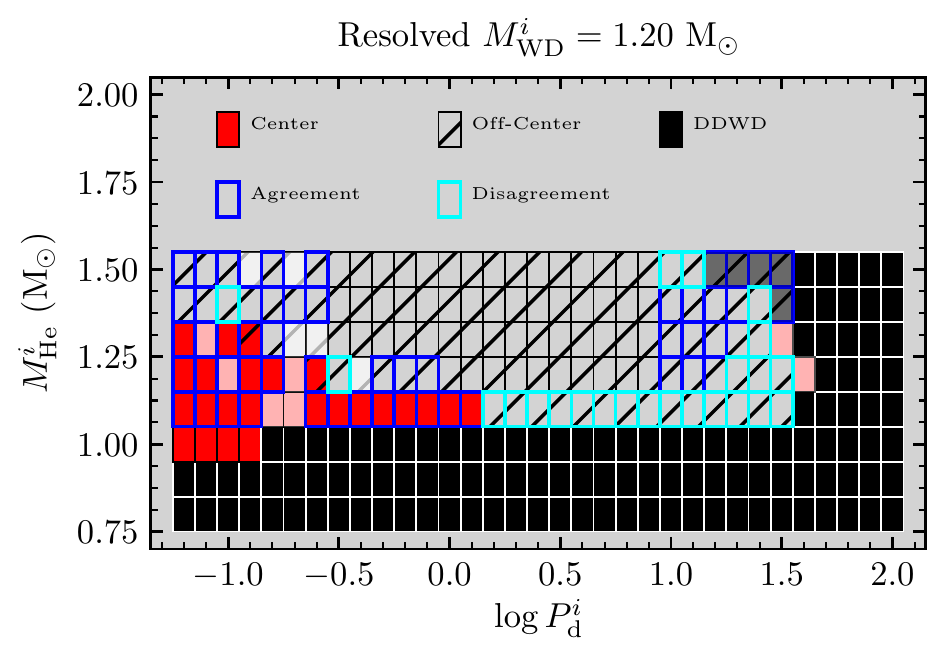}{0.5\textwidth}{(b)}
}
\caption{Final outcome as a function of initial binary parameters using point-mass WD models (panel a) and, for selected systems (shown with blue edges), resolved WD models (panel b). Red boxes represent systems with central ignitions, boxes with black stripes show systems with off-center ignitions, and black boxes represent detached double WD binaries. In Panel b, dark (light) blue edges represent agreement (disagreement) in the final outcomes between the point-mass and resolved models.
}
\label{fig:resolved_m1p20}
\end{figure*}

We first create a grid of binary models with a point-mass WD using the
approach described in Section~\ref{sec:preexp-evol} and using the helium
flash retention efficiencies of \Kato.  The resulting outcomes are
shown in Panel (a) of Figure~\ref{fig:resolved_m1p20}. The black boxes
identify systems that form detached double WD binaries.  In the other
systems, the WD eventually reaches $\mex$ $(\approx \mch)$.  Based on the
results of resolved CO WD models accreting at constant rates,
\citet{Wang2017i} propose that off-center carbon ignitions can be
approximately detected by comparing the final mass transfer rate to a critical value,
$\dot{M}^{f}_{\mathrm{WD}} \ge \mdotcr = \unit[2.05 \times 10^{-6}]{\msunyr}$.
Systems that satisfy this criterion are marked by
grey boxes with black stripes.  The remaining systems that do not
satisfy the \citet{Wang2017i} criterion are then designated as
Chandrasekhar-mass central ignitions and indicated by red boxes.

Similar to the results of \citet{Wang2017i}, the point-mass grid suggests that an initially 1.20 $\msun$ CO WD may undergo central ignition for an orbital period up to $\approx 30 $ days. These long period systems may then result in cool pre-explosion He stars. 

\subsection{Resolved WD binary models}

We resimulate a portion of the point-mass grid, using a resolved WD model in order to account for off-center carbon ignitions self-consistently. For these models, we construct a 1.20 $\msun$ CO WD in $\mesa$ by scaling up the mass of a $1.00 \, \msun$ CO WD. We evolve the binary until we identify one of the two following outcomes:

% The resimulation results are plotted over the point-mass grid in Panel (b) of Figure~\ref{fig:resolved_m1p20}, where the resimulated models are shown with dark blue or light blue edges.

\begin{enumerate}

\item \textbf{Off-center ignition.} We identify the models with
  off-center ignitions in the WD as those where energy release rate from
  carbon burning ($\epsilon_{\mathrm{CC}}$) exceeds the non-nuclear
  neutrino cooling rate ($\epsilon_{\nu}$) at an off-center temperature peak.

\item \textbf{Central Ignition.} We identify the models with central
  ignitions in two ways.  When central ignition of the WD occurs
  during stable accretion, we directly see
  $\epsilon_{\mathrm{CC}} \ge \epsilon_{\nu}$ at the WD center during the calculation.  For models where helium flashes
  begin following stable
  accretion, we stop the calculation after a few flashes.  If the corresponding point-mass model indicates that
  the system will produce a near $\mch$ WD, then we also classify it as central ignition.
\end{enumerate}

Panel (b) of Figure \ref{fig:resolved_m1p20} summarizes the results.  To facilitate comparison,
models that agree with the outcome in panel (a) are shown with dark blue edges, while models that disagree are shown with light blue edges.
Some of the models which we attempted to resimulate failed due to computational difficulties associated with He flashes on the WD and so the outcome is indeterminate. These systems are masked by a white box in Panel (b), resulting in pink boxes (for central ignitions) and light grey boxes with black stripes (for off-center ignitions).

In contrast to the method of \cite{Wang2017i} as applied in panel (a),
our resolved WD models in panel (b) show off-center carbon ignitions
for long period systems ($\iniP \ge 0.2 $). This is because for the
long period systems, the mass transfer rate has decreased
significantly by the time WD approaches $\mch$, leading to $|\mdothe|$
below the value of $\mdotcr$, and thus identification
as central ignitions according to the criterion of \citet{Wang2017i}.
However, earlier in the
evolution, well before the WD approaches $\mch$, an off-center
ignition already occurred during a phase with higher $\mdothe$.

\begin{figure}
  \centering
  \includegraphics[width=\columnwidth]{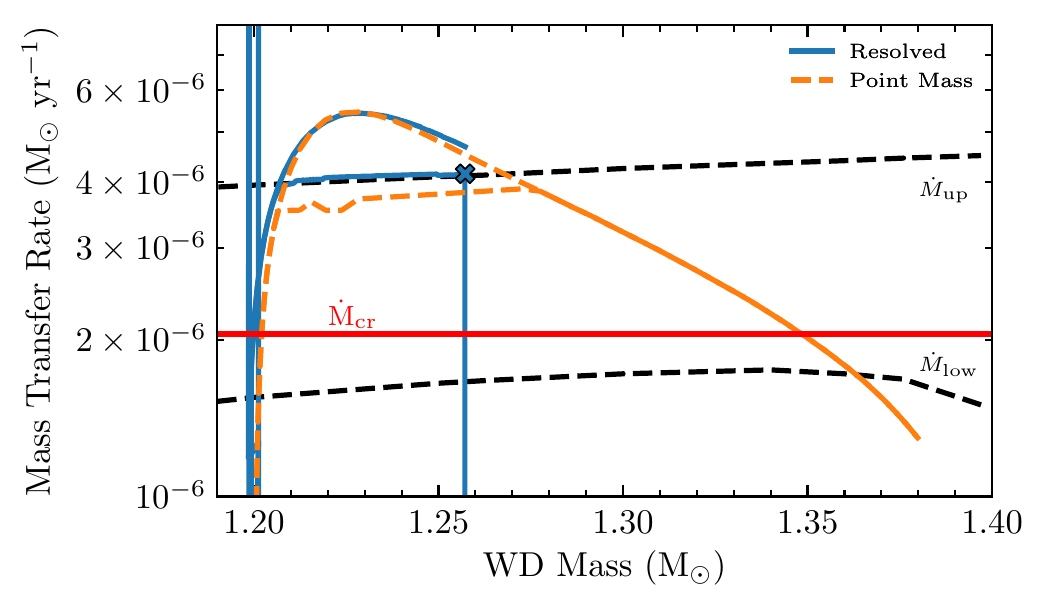}
  \caption{Mass transfer histories of two systems with $\iniMwd=1.20\, \msun$ and $\iniMhe=1.1\, \msun$, at an initial orbital period of $\iniP = 1.0$. In the point mass models (dashed), the WD reaches $\mch$.  The mass transfer rate has declined significantly from peak, reaching $\mdotwd < \mdotcr$, and so the method of \cite{Wang2017i} suggests a central ignition in the WD. However, in the resolved models, off-center ignition occurs early on. The difference between the mass transfer histories in their region of overlap is due to the fact that the point mass model uses a fitted version of $\mdotup$ which is $\approx 10\%$ lower than one realized in the resolved model and because of the initial He flashes that occur in the resolved model.}
  \label{fig:mdot_history_m1p20}
\end{figure}

This is demonstrated by the mass transfer history in Figure
\ref{fig:mdot_history_m1p20}.  The mass transfer peaks above
$\mdotup$, and so the WD accretes at this roughly constant rate of
$\approx \unit[4\times10^{-6}]{\msunyr}$.  As found by
\citet{Wang2017i}, a WD accreting at this constant rate (which is
$>\mdotcr$) experiences an off-center ignition.  Here, that happens
after the WD has grown to $\mwd \approx \unit[1.25]{\Msun}$.
However, if the evolution is allowed to continue (as in the point
mass calculation) the accretion rate falls.  By the time the WD
reaches $\unit[1.38]{\Msun}$, the accretion rate has fallen below
$\mdotcr$ and so the prescription of \citet{Wang2017i}--which
considers $\dot{M}$ at only this final point--classifies this as a
central ignition.  Because the off-center ignition occurs after only
accreting a relatively small amount of mass ($\approx 0.05\,\Msun$),
detecting its occurrence requires a prescription that accounts for the
changing mass transfer rate throughout the evolution.

\subsection{Results and Implications}

\begin{figure*}
\gridline{
\fig{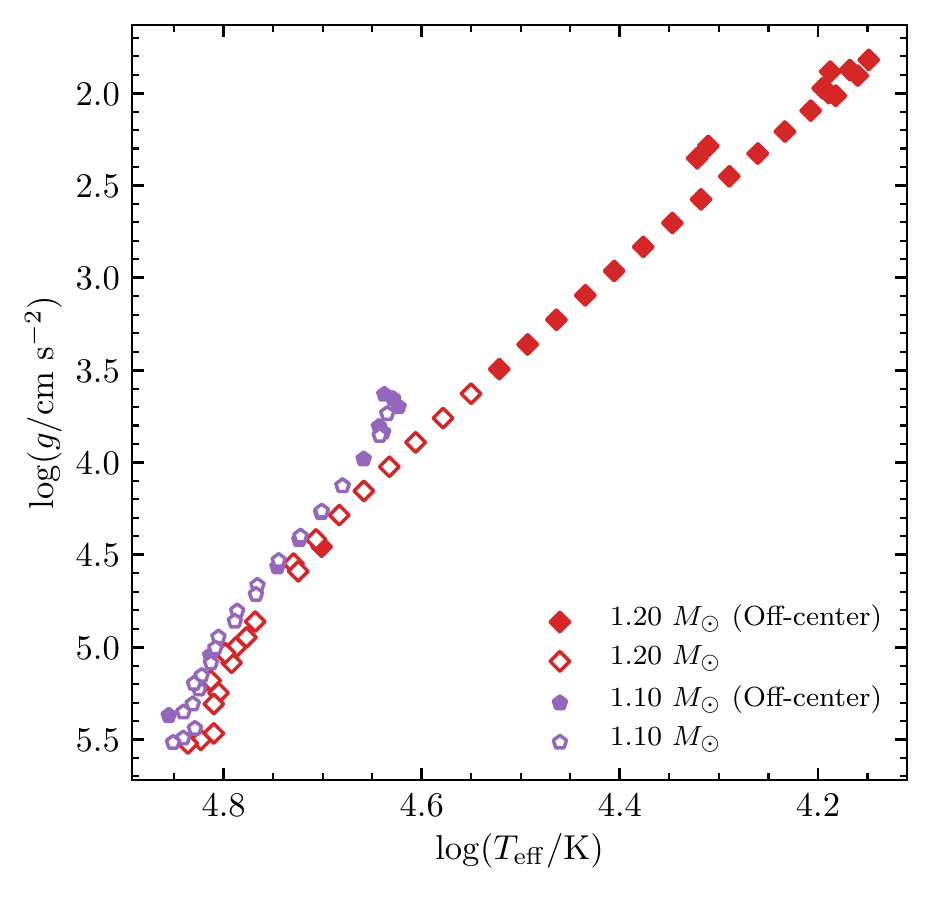}{0.5\textwidth}{(a)}
\fig{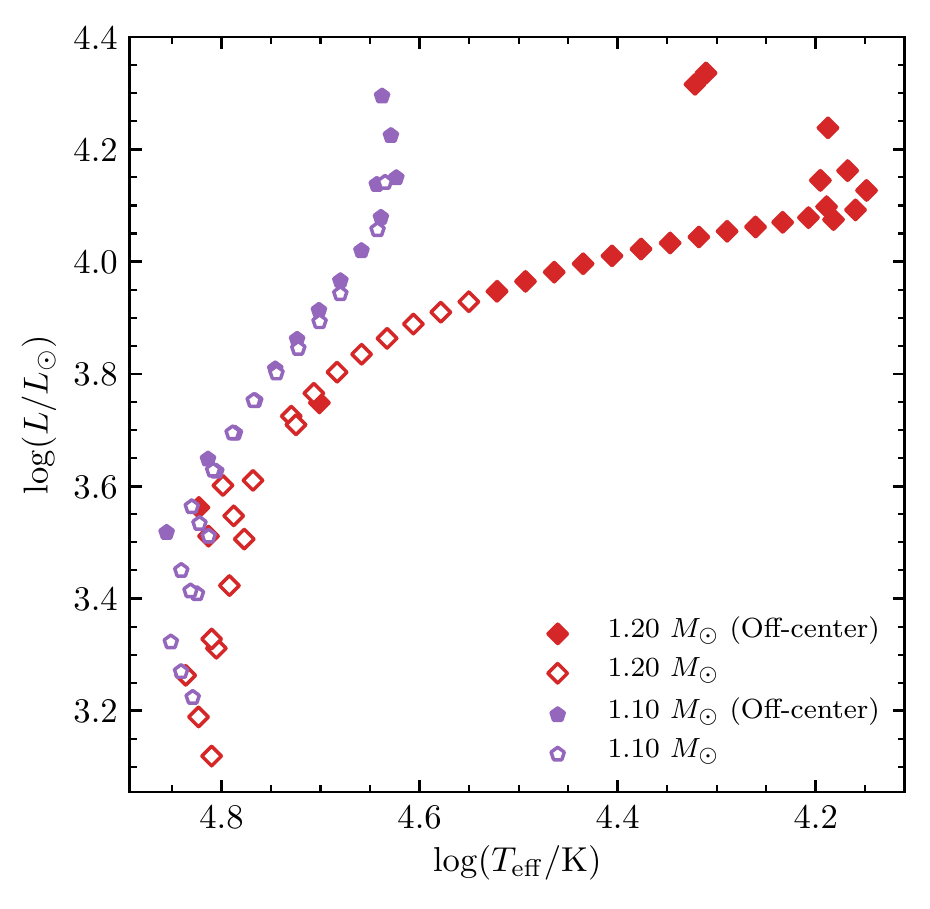}{0.5\textwidth}{(b)}
          }
\caption{
  Kiel diagram (panel a) and Hertzsprung-Russell diagram (panel b) for pre-explosion He star models with ultra-massive CO WDs.
  Unfilled symbols indicate central ignitions, 
  while filled symbols indicate models that undergo off-center ignitions and will not explode as TN SNe.
  Other aspects are the same as Figure~\ref{fig:logg-logTeff}. 
  \label{fig:logg-logTeff_highmass}
}
\end{figure*}

\begin{figure*}
    \fig{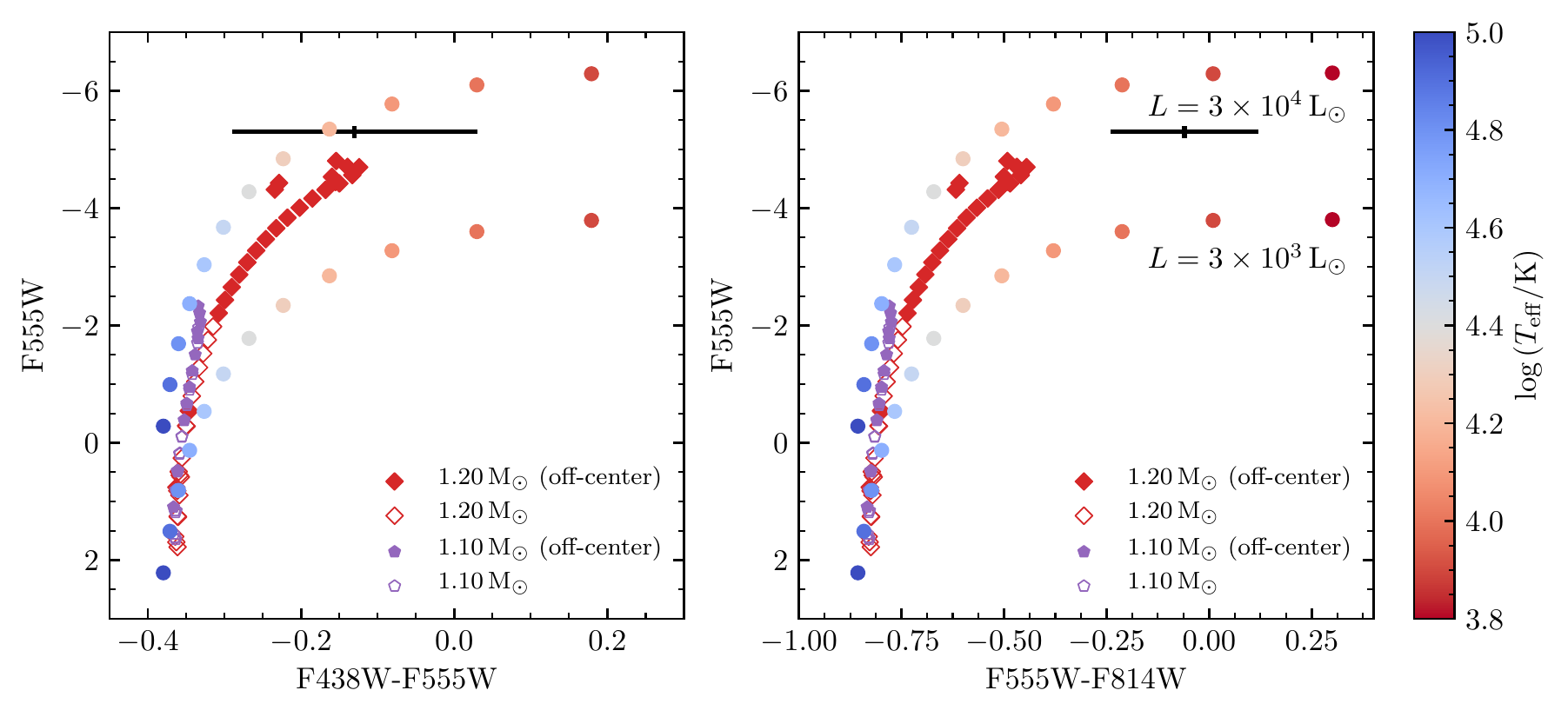}{\textwidth}{}
    \caption{Color-magnitude diagrams for indicated optical WFC3/UVIS filters for ultra-massive CO WD models.
      Unfilled symbols indicate central ignitions, 
      while filled symbols indicate models that undergo off-center ignitions and will not explode as TN SNe.
      Error bars indicate the pre-explosion source observed for SN 2012Z by \citet{McCully2014}.
    \label{fig:colors_highmass}}
\end{figure*}

We similarly run another set of resolved models starting with a 1.10 $\msun$ CO WD.
Figure~\ref{fig:logg-logTeff_highmass} places the models for both
WD masses on the Kiel
and HR diagrams; Figure~\ref{fig:colors_highmass} shows them on
color-magnitude diagrams.  In each plot, the filled points indicate
the models that we identify as undergoing off-center ignition, and
hence not undergoing a TN SN, but that previous work would have
identified as having done so.  Eliminating the off-center ignitions
serves to eliminate the coolest and most luminous He star companions.

The fact that our resolved 1.20 $\msun$ CO WD models in long-period binaries undergo off-center ignitions has significant implications for our understanding the progenitor of SN 2012Z. If SN 2012Z-S1 is a He star - WD binary, then either (i) SN 2012Z originates from a long-period ($\gtrsim 10$ days) He star - WD binary where the initial WD is massive ($\iniMwd \approx 1.2 \, \msun$) but not a CO WD (so possibly a hybrid CO/ONe WD or an ONe WD), or (ii) the pre-explosion optical light from the system is not dominated by the (unmodified) emission from the He star.

\section{Discussion and Conclusions}
\label{sec:conclusions}

In this work, we have further investigated the He star - CO WD
progenitor channel for thermonuclear supernovae, with a particular
focus on the predicted properties of the donor He star at the time the
WD explodes.  In Section~\ref{sec:preexp-evol}, we describe an
extension of the binary evolution calculations of \citet{Wong2019}
that allowed us to generate a set of He star models at the time the WD
explodes over a large range of initial binary parameters.  In
Section~\ref{sec:preexp-models}, we characterize the pre-explosion
properties of the donor stars.  Using stellar atmosphere models, we
demonstrated that the blackbody assumption is sufficient for
characterizing the optical emission from these stars.  We compared the
optical emission from our models to the properties of the source
observed in pre-explosion imaging of the Type Iax SN 2012Z.  In
agreement with past work, we found that binaries with normal
$(\lesssim \unit[1.05]{\Msun})$ CO WDs dramatically fail to reproduce
these observations. In Section~\ref{sec:massive}, we made models
beginning with ultra-massive $(\approx \unit[1.2]{\Msun})$ CO WDs.  If the
WD is approximated as a point mass, such models have been previously
demonstrated to better match the properties of the 2012Z pre-explosion
imaging.  However, our models, which resolved the internal structure
of the CO WD accretor, show that such systems undergo off-center
carbon ignition and thus are not expected to produce
thermonuclear supernovae.

We therefore conclude that, under the assumption that the He star
donor dominates the optical light of the system, our self-consistent
He star - CO WD binary models fail to reproduce the properties of the
detected source in pre-explosion imaging of the host galaxy of SN
2012Z \citep{McCully2014}. The other Type Iax SN with similarly deep pre-explosion host observations is SN 2014dt \citep{Foley2015a}.  That case resulted in a non-detection, as did SN 2008ge \citep{Foley2010} which has shallower limits, and so both are consistent with our models. 

The motivation for invoking the Chandrasekhar-mass, He star donor channel for Type Iax SNe remains \citep[e.g.,][]{Jha2017}.
The presence of strong Ni emission in the late-time spectra \citep{Foley2016} suggests the high density explosion characteristic of a near-$M_{\rm Ch}$ WD.  Population synthesis studies have shown that the He star channel contributes to TN SNe with delay times $\lesssim 100 $ Myr \citep{Wang2009b, Claeys2014}, consistent with the typical delay time of $\approx 60 $ Myr of SNe Iax, inferred from their nearby stellar populations \citep{Takaro2020}.  In this study, we classified of the final outcome based on the location of carbon ignition in the accreting WD.  Open questions remain about the evolution beyond the phase of off-center carbon ignition in massive WDs \citep[e.g.,][]{Wu2019, Wu2020}, so it is possible that further progress will revise our understanding of which systems can explode, thereby altering the predicted companion properties.

Additionally, the fact that these are binary systems with complex evolutionary histories is not fully addressed by only considering the He star.
Both the WD and its accretion disk can also be luminous, though the expected high effective temperatures of this emission imply
these sources are sub-dominant in the optical.
Material has also likely been ejected into the circumstellar environment
due to non-conservative mass transfer and He novae.%
\footnote{In a recent study, \citet{Moriya2019} consider this environment and show the expected circumstellar density is consistent with the non-detection of radio emission in a number of observed events.}
This may be able to modify the emission: significant circumstellar reddening from
carbon-rich ejecta has been invoked for the He nova V445 Pup, see
\citealt{Woudt2009}).  Developing a more detailed understanding of the
combined influence of the binary and its environment  will be an important avenue for future work.

% \sunny{From blackbody modeling the accreting WD would be $\sim 3 $ orders of magnitude fainter than the He star in the optical. However the accretion disk would be comparable to the He star (at least the short-period He stars).  }

% The pre-explosion colors may be affected by two factors -- one, the
% presence of an accreting WD; two,

This investigation of this progenitor channel will be aided by other complementary probes. 
The He-star donor is Roche-lobe filling at the
time of explosion and so material from its outer layers may be
entrained due the impact of SN ejecta.  Thus this He star donor
scenario can also be constrained by limits on the inferred amount of He
present in late time spectra. Our models predict a remaining He envelope mass of $M_{\text{He env}} \approx 0.06 - 0.55 \, \msun$. Current theoretical models predict a
stripped He mass $\sim \unit[10^{-2}]{\msun}$, assuming a typical He star
radius of $\Rhe \approx 0.5 \, \rsun$ \citep{Liu2013}, in tension with
emerging observational limits of $\lesssim \unit[10^{-3}]{\msun}$
\citep{Magee2019} and $\lesssim \unit[10^{-2}]{\msun}$  \citep{JacobsonGalan2019}. However, a stripped mass of $ \approx 3 \times 10^{-4} \, \msun$ was found in the simulation of \cite{Zeng2020} who assumed a weak pure deflagration model. 
We also note that a limit of $\lesssim 2 \times 10^{-3} \, \msun$ is found for stripped H in SNe Iax \citep{JacobsonGalan2019}.

Nevertheless, even in the absence of significant stripping, the He star donor channel may provide an explanation for the detection of helium emission in the early-time spectra of SNe 2004cs and 2007J \citep{JacobsonGalan2019}. In some systems of our simulated grid, the pre-explosion mass-transfer rate drops below the stable regime, such that the WD undergoes He novae and ejects He-rich material into the environment. This is consistent with the inference of the He emission to be originating from circumstellar He \citep{JacobsonGalan2019}.

Searching for the surviving He star companion may offer another constraint on the He star donor channel. If the surviving companion is relatively unperturbed, a search in the UV may be useful, for nearby TN SNe (see our Figure \ref{fig:colors}). However, our pre-explosion He star models appear to be too blue for the post-explosion source found in SN 2008ha \citep{Foley2014} and in SN 2012Z \citep{McCully2021}. On the other hand, the simulations by \cite{Pan2013} show that the surviving companion brightens significantly, and may alter its colors, in a timescale in $\approx 10-30$ years. Their binaries are much more compact than ours, with their longest period binary at 3610 s ($\finalP \approx -1.40$), so it is unclear how our He stars in wider binaries are impacted by the SN ejecta. Future simulations of He star-ejecta interaction in a long-period system, with various degrees of He envelope-stripping, may help shed light on this problem.

\begin{acknowledgments}
We thank Ryan Foley, Wolfgang Kerzendorf, Enrico Ramirez-Ruiz, Silvia Toonen, and Stan Woosley for helpful conversations. We thank the anonymous referee for their constructive comments that have improved this manuscript. 
J.S. is supported by the National Science Foundation through grant ACI-1663688.
Support for this work was provided by NASA through Hubble Fellowship
grant \# HST-HF2-51382.001-A awarded by the Space Telescope Science
Institute, which is operated by the Association of Universities for
Research in Astronomy, Inc., for NASA, under contract NAS5-26555. 
The simulations were run on the Hyades supercomputer at UCSC, purchased
using an NSF MRI grant. 
Use was made of computational facilities purchased with funds from the National Science Foundation (CNS-1725797) and administered by the Center for Scientific Computing (CSC). The CSC is supported by the California NanoSystems Institute and the Materials Research Science and Engineering Center (MRSEC; NSF DMR 1720256) at UC Santa Barbara.
This research made extensive use of NASA's
Astrophysics Data System.
\end{acknowledgments}

% ============================================================

\software{%
  \texttt{MESA} \citep[v10398;][]{Paxton2011, Paxton2013, Paxton2015, Paxton2018, Paxton2019},
  \texttt{ipython/jupyter} \citep{perez_2007_aa,kluyver_2016_aa},
  \texttt{matplotlib} \citep{hunter_2007_aa},
  \texttt{NumPy} \citep{der_walt_2011_aa}, and
  \texttt{starkit} (\url{https://github.com/starkit/starkit}),
  \texttt{wsynphot} (\url{https://github.com/starkit/wsynphot}),
  \texttt{Python} from \href{https://www.python.org}{python.org}.
}

\clearpage

\appendix

\twocolumngrid

\section{Modeling the Donor Mass Transfer}
\label{sec:appendixA}

During this work, we became aware that some of our He star models, particularly the long-period systems with $\iniMwd = 0.90$ and 0.95 $\msun$, have radii that exceed their Roche lobe radii by up to factors of a few. 
We believe that this behavior is unphysical because the mass transfer rate is expected to increase exponentially as the donor overfills its Roche lobe \citep[e.g.,][]{Ritter1988}. 
The large overfill factors are seen when the He star starts to come out of contact and the envelope becomes highly radiation-dominated. 
Under these conditions, us adopting $\tt tau\_factor=100$ (i.e., the surface cell is placed at an optical depth of $\tau = 100 \times 2/3 $) appears to have an effect on the envelope structure. 
In this Appendix, we test the effects of adopting different physical assumptions, namely adopting the \texttt{Kolb} mass transfer scheme \citep{Kolb1990} which considers optically-thick mass transfer, and setting $\tt tau\_factor=1$ for the He star. With these two changes, we re-simulated all the point-mass models shown in this work, with $\iniMwd$ ranging from $0.9$ to $1.2 \, \msun$. 

\begin{figure*}
\gridline{
\fig{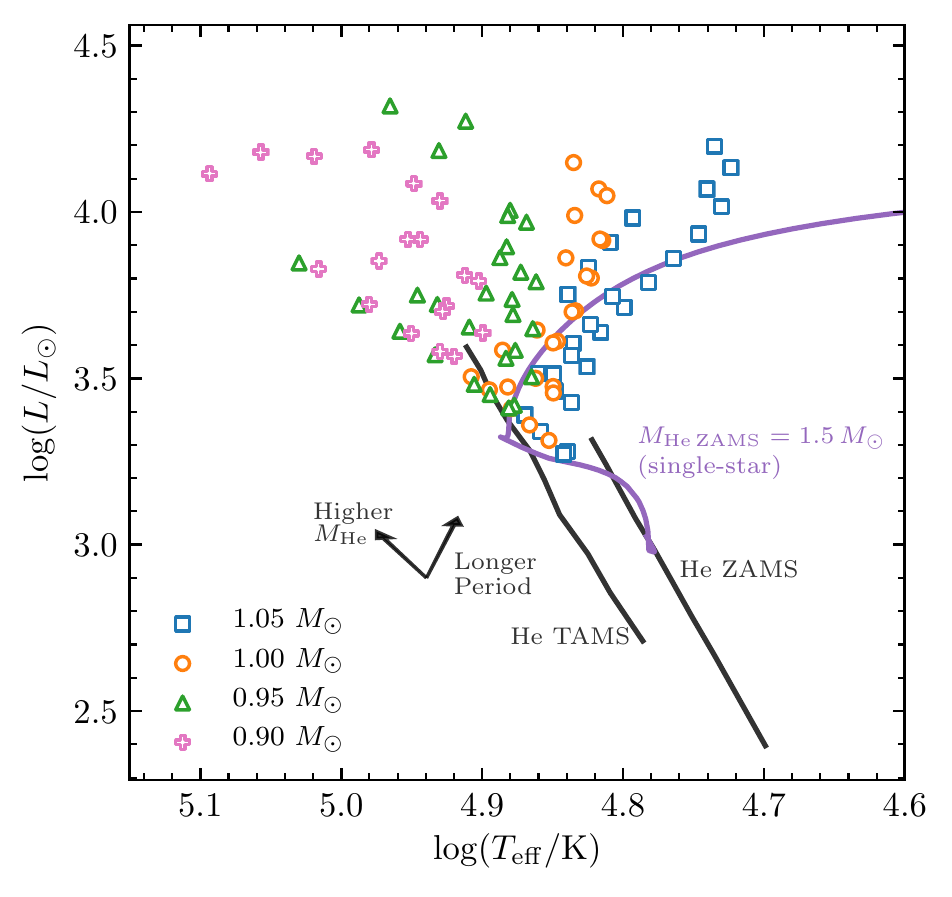}{0.5\textwidth}{(a)}
\fig{combined_HR_highMass_Kolb.pdf}{0.5\textwidth}{(b)}
          }
\caption{
  Hertzsprung-Russel diagrams for pre-explosion He-star models with $\iniMwd=0.9-1.05 \, \msun$ CO WDs (panel a), and with $\iniMwd=1.1-1.2 \, \msun$ ultra-massive CO WDs (panel b), same as Figures \ref{fig:logg-logTeff} and \ref{fig:logg-logTeff_highmass} but with \texttt{Kolb} and $\tt tau\_factor=1$ adopted for the models. 
  \label{fig:HR_Kolb}
}
\end{figure*}

The resulting HR diagrams for the pre-explosion He stars are shown in Figure \ref{fig:HR_Kolb}. For $\iniMwd = 0.9 - 1.05 \, \msun$ (panel a), adopting \texttt{Kolb} and $\tt tau\_factor=1$ in general increases the mass retention efficiency, and allows a few more systems on the boundary of the TN SN region to reach $\mch$. It also keeps the He star radii to within $\approx 10 \%$ of the Roche lobe radii for systems in the TN SN region. As a result, the pre-explosion He stars have higher $\Teff$ compared with Figure \ref{fig:logg-logTeff}, panel b. This change in $\Teff$ mainly affects the long-period systems in which the He star has nearly exhausted its envelope. We also note that the mass transfer history near peak $\mdothe$ is relatively unchanged. As expected, we only see a change in mass transfer history for the long-period systems as the donor starts to come out of contact. 

Similarly, the mass transfer histories show good agreement between the new $\tt tau\_factor=1$ and the old $\tt tau\_factor=100$ point-mass runs, for the $\iniMwd = 1.1 \, \msun $ models, and the $\iniMwd = 1.2 \, \msun $ models with $\iniP \leqslant 0.2 - 0.3 $. It is therefore not surprising that for these models, the TN SN regions obtained by applying the $\mdotcr$ criterion of \cite{Wang2017i} remain nearly unchanged.
%(except for $\inipara = (1.1,1.2,0.0)$). 

However, the mass transfer history starts to differ for $\iniMwd = 1.2 \, \msun $ and $\iniP \gtrsim 0.3 $,  because of a surface convection zone that was not captured previously with $\tt tau\_factor=100$. With longer $\iniP$, the surface convection zone encloses more mass and increasingly changes the behavior of $\mdothe$ near peak. Peak $\mdothe$ increases, so that the He star exhausts its envelope more easily. While the outcomes for $\iniMwd = 1.2 \, \msun $ and $\iniP \leqslant 0.5 $ remain unchanged, a discrepancy in outcome starts to arise for $\iniP > 0.5 $, and in turn the TN SN boundary moves to shorter $\iniP$. 

We do not re-simulate resolved models for $\iniMwd = 1.1 $ and $1.2 \, \msun$, and thus cannot identify with certitude systems that the $\mdotcr$ criterion of \cite{Wang2017i} would misclassify as one that undergoes a central ignition. However, for $\iniMwd = 1.1 \, \msun $, and $\iniMwd = 1.2 \, \msun $ with $\iniP \leqslant 0.2 - 0.3 $, given the good agreement in mass transfer histories, we can still rely on our prior identification. Furthermore, our resolved $\iniMwd = 1.2 \, \msun $ models invariably undergo off-center ignition if the WD is always accreting at $\mdotwd = \mdotup$ until its mass grows to $\mwd \approx 1.26 \, \msun$ (see also Figure \ref{fig:mdot_history_m1p20}). This is always the case for the systems with $\iniMwd = 1.2 \, \msun $ and $\iniP \gtrsim 0.3 $ that reach $\mch$, so we can also reliably classify them as systems would undergo off-center ignition. 

The HR diagram for $\iniMwd = 1.1 $ and $1.2 \, \msun$ is shown in in Figure \ref{fig:HR_Kolb}, panel b. Compared with Figure \ref{fig:logg-logTeff_highmass}, panel b, the locations of $\iniMwd = 1.1 $ systems show little change, and as do the $\iniMwd = 1.2 $ systems that we expect would experience a central ignition. The $\iniMwd = 1.2 $ systems that we expect would experience an off-center ignition are located at slightly higher $\Teff$, due to the change in mass transfer history for $\iniP > 0.2 - 0.3 $. 

Overall, these pre-explosion He star models with different assumptions about the mass transfer and the envelope of the He star show similar properties to the models we presented in the main body of this work. Therefore, our conclusions remain unchanged. However, this Appendix together with the necessity for us to use the \texttt{MLT++} capacity of $\mesa$ serve to highlight the uncertainties associated with modeling mass transfer from a donor star with a highly radiation-dominated envelope, which remains a caveat of our work.

\clearpage

\bibliographystyle{aasjournal}
\bibliography{preexp,references_bin.bib}

\end{document}